\documentclass{aa}
\usepackage{graphicx}
\usepackage{bm}
\usepackage{times}
\usepackage{amssymb}
\graphicspath{{./fig/}{./png/}}
\newcommand{\EQ}{\begin{equation}}
\newcommand{\EE}{\end{equation}}
\newcommand{\EQA}{\begin{eqnarray}}
\newcommand{\EEA}{\end{eqnarray}}
\newcommand{\brac}[1]{\langle #1 \rangle}
\newcommand{\pd}{\partial}

\newcommand{\ve}[1]{\boldsymbol{#1}}
\newcommand{\mean}[1]{\overline{#1}}
\newcommand{\meanv}[1]{\overline{\bm #1}}
\newcommand{\cst}{c_{\rm s}^2}

\newcommand{\urms}{u_{\rm rms}}

\newcommand{\Ma}{{\rm Ma}}

\newcommand{\ru}{\hat{\bm r}}

\newcommand{\kef}{k_{\rm f}}
\newcommand{\chit}{\chi_{\rm t}}
\newcommand{\chitz}{\chi_{\rm t0}}
\newcommand{\tauc}{\tau_{\rm c}}

\newcommand{\St}{{\rm St}}

\newcommand{\Pra}{{\rm Pr}}
\newcommand{\Ra}{{\rm Ra}}

\newcommand{\Rey}{{\rm Re}}
\newcommand{\Co}{{\rm Co}}

\newcommand{\qij}{Q_{ij}}
\newcommand{\qrt}{Q_{r\theta}}
\newcommand{\qrp}{Q_{r\phi}}
\newcommand{\qtp}{Q_{\theta\phi}}

\newcommand{\nab}{\mbox{\boldmath $\nabla$} {}}
\newcommand{\OOO}{\hat{\mbox{\boldmath $\Omega$}} {}}
{}
\def\onethird{{\textstyle{1\over3}}}
\def\onehalf{{\textstyle{1\over2}}}
\def\threehalfs{{\textstyle{3\over2}}}

\newcommand{\Fig}[1]{Fig.~\ref{#1}}

\begin{document}

\authorrunning{K\"apyl\"a et al.}
\titlerunning{Reynolds stress and heat flux in spherical shell convection}

   \title{Reynolds stress and heat flux in spherical shell convection
}

   \author{P. J. K\"apyl\"a
          \inst{1,2}
          \and
          M. J. Mantere
          \inst{1}
           \and 
         G. Guerrero
          \inst{2}
          \and
          A. Brandenburg
          \inst{2,3}
          \and
          P. Chatterjee
          \inst{2}
          }

   \offprints{\email{petri.kapyla@helsinki.fi}
              }

   \institute{Department of Physics, Gustaf H\"allstr\"omin katu 2a 
              (PO Box 64), FI-00014 University of Helsinki, Finland
         \and NORDITA, Roslagstullsbacken
              23, SE-10691 Stockholm, Sweden
         \and Department of Astronomy, Stockholm University, SE-10691
              Stockholm, Sweden}

   \date{Received 6 October 2010 / Accepted 22 April 2011}

   \abstract{Turbulent fluxes of angular momentum and enthalpy or heat due to
     rotationally affected convection play a key role in determining
     differential rotation of stars.
     Their dependence on latitude and depth has been determined in the past
     from convection simulations in Cartesian or spherical simulations.
     Here we perform a systematic comparison between the two geometries
     as a function of the rotation rate.
   }{%
     Here
     we want to extend the earlier studies by using
     spherical wedges to obtain turbulent angular momentum and heat transport as
     functions of the rotation rate from stratified 
     convection. We compare results from spherical and Cartesian
     models in the same parameter regime in order to study whether 
     restricted geometry introduces artefacts into the results.
     In particular, we want to clarify whether the sharp equatorial profile
     of the horizontal Reynolds stress found in earlier Cartesian models is
     also reproduced in spherical geometry.
   }%
   {We employ direct numerical simulations of turbulent convection in
     spherical and Cartesian geometries. In order to alleviate the 
     computational cost in the spherical runs,
     and to reach as high spatial resolution as possible, we model
     only parts of the latitude and longitude. The rotational
     influence, measured by the Coriolis number or inverse Rossby
     number, is varied from zero to roughly seven, which is the regime that is
     likely to be realised in the solar convection zone. Cartesian
     simulations are performed in overlapping parameter regimes.
   }%
   {For slow rotation we find that the radial and latitudinal
     turbulent angular momentum fluxes are directed inward and
     equatorward, respectively. In the rapid rotation regime the
     radial flux changes sign in accordance with earlier numerical
     results, but in contradiction with theory. The latitudinal flux
     remains mostly equatorward and develops a maximum close to the
     equator. In Cartesian simulations this peak can be explained by
     the strong `banana cells'. Their effect in the spherical case
     does not appear to be as large. The latitudinal heat flux is
     mostly equatorward for slow rotation but changes sign for rapid
     rotation. Longitudinal heat flux is always in the retrograde
     direction. The rotation profiles vary from anti-solar (slow
     equator) for slow and intermediate rotation to solar-like (fast
     equator) for rapid rotation. The solar-like profiles are dominated by the
     Taylor--Proudman balance.}%
   {}%


   \maketitle

%____________________________________________________________

\section{Introduction}
The surface of the Sun rotates differentially: the rotation period at
the poles is roughly 35 days as opposed to 26 days at the
equator. Furthermore, the internal rotation of the Sun has been
revealed by helioseismology (e.g.\ Thompson et al.\
\cite{Tea03}): the radial gradient of $\Omega$ is small in the bulk of
the convection zone, whereas regions of strong radial differential
rotation are found near the base and near the surface of the convection zone. 
According to dynamo theory, large-scale shear plays an important
role in generating large-scale magnetic fields (e.g., Moffatt
\cite{M78}; Krause \& R\"adler \cite{KR80}).
More specifically, large-scale shear lowers the threshold for dynamo
action and the combined effect of helical turbulence and shear yields
oscillatory large-scale magnetic fields, resembling the observed solar
activity pattern (e.g.\ Yoshimura \cite{Y75}). It is even possible to
drive a large-scale dynamo in nonhelical turbulence with shear (e.g.,
Brandenburg \cite{B05}; Yousef et al.\ \cite{Yea08a,Yea08b};
Brandenburg et al.\ \cite{BRRK08}). Thus, it is of great interest to
study the processes that generate large-scale shear in solar and
stellar convection zones.

Differential rotation of the Sun and other stars is thought to be
maintained by rotationally influenced turbulence in their
convection zones. In hydrodynamic mean-field theories of stellar 
interiors the
effects of turbulence appear in the form of turbulent fluxes of angular
momentum and enthalpy or heat (cf.\ R\"udiger \cite{R89}; R\"udiger \& Hollerbach
\cite{RH04}). 
These fluxes can be defined by Reynolds averaging of
products of fluctuating quantities, v.i.z.,  
the fluxes of angular momentum and heat, respectively, are
\begin{eqnarray}
Q_{ij}&=&\mean{u_i' u_j'},\\
F_{i}&=&c_{\rm P} \mean{\rho}\;\mean{u_i' T'}.
\end{eqnarray}
Here overbars denote azimuthal averaging,
primes denote fluctuations about the averages,
$Q_{ij}$ is the Reynolds stress,
$F_i$ is the turbulent convective energy flux,
$\bm{u}$ is the velocity, $T$ is the temperature, $\rho$ is density,
and $c_{\rm P}$ is the specific heat at constant pressure.

Much effort has been put into computing these correlations using
analytical theories (e.g., R\"udiger \cite{R80,R82}; 
Kitchatinov \& R\"udiger \cite{KR93}; Kitchatinov et al.\ 
\cite{KPR94}). Most of the
analytical studies, however, rely on approximations such as
first-order smoothing, the applicability of which in the stellar environments
can be contested. In order to get more insight, idealised numerical
simulations, often working in Cartesian geometry, have been
extensively used to compute the stresses for modestly large Reynolds
numbers (e.g., Pulkkinen et al.\ \cite{Pea93}; Brummell et al.\ 
\cite{BHT98}; Chan \cite{C01};
K\"apyl\"a et al.\ \cite{KKT04}; R\"udiger et al.\
\cite{R05b}). However, the Cartesian simulations have yielded some
puzzling results, such as the latitudinal angular momentum flux having a very 
strong maximum
very close to the equator (e.g., Chan \cite{C01}; Hupfer et al.\
\cite{HKS05}) and a sign change of the corresponding radial flux
(K\"apyl\"a et al.\ \cite{KKT04}). Neither of these effects can be
recovered from theoretical studies (R\"udiger \& Hollerbach
\cite{RH04}) or simpler forced turbulence
simulations (K\"apyl\"a \& Brandenburg \cite{KB08}).
The Reynolds stresses have also been computed from high resolution 
spherical convection simulations (e.g.\ DeRosa et al.\ \cite{DGT02}; 
Miesch et al.\ \cite{MBDRT08}), but a detailed comparison with 
Cartesian results is lacking in the literature.

Rotation also affects the turbulent convective energy transport.
In fact, in the presence of rotation, the turbulent heat transport due to
convection is no longer purely radial (e.g.,
Brandenburg et al.\ \cite{BMT92}; Kitchatinov et al.\ 
\cite{KPR94}; Brun \& Rempel \cite{BR09}).
In a sphere, such anisotropic heat transport
leads to latitude-dependent temperature and entropy distributions.
Such variations can be important in determining the rotation profile of 
the Sun: neglecting the Reynolds stress and molecular diffusion, the 
evolution of the azimuthal component of vorticity,
$\meanv\omega=\bm\nabla\times\meanv{u}$, is governed by
\begin{equation}
\frac{\pd \mean{\omega}_\phi}{\pd t} = r \sin \theta \frac{\pd \mean{\Omega}^2}{\pd z} + \frac{1}{\mean{\rho}^2} (\bm\nabla\mean{\rho} \times \bm\nabla \mean{p})_\phi,\label{equ:wphi}
\end{equation}
where $\pd/\pd z=\OOO\cdot\nab$ is the derivative along the unit vector
of the rotation vector, $\OOO=(\cos\theta,-\sin\theta,0)$,
and $p$ is the pressure.
The last term on the rhs
describes the baroclinic term which can be written as
\begin{equation}
\frac{1}{\mean{\rho}^2}
(\bm\nabla\mean{\rho} \times \bm\nabla \mean{p})_\phi
=(\bm\nabla\mean{T} \times \bm\nabla \mean{s})_\phi
\approx-\frac{g}{rc_{\rm P}}\frac{\pd \mean{s}}{\pd \theta}.
\end{equation}
where $g=|\vec{g}|$ is the acceleration due to gravity,
$s$ is the specific entropy, and
$\bm\nabla\mean{T}\approx\vec{g}/c_{\rm P}$ has been used
for the adiabatic temperature gradient.
In the absence of latitudinal entropy gradients, the solution of
Eq.~(\ref{equ:wphi}) is given by the Taylor--Proudman theorem, i.e.\
$\pd\mean{\Omega}/\pd z=0$. In general, however, the thermodynamics
cannot be neglected and latitudinal
gradients of entropy influence the rotation profile of the star via
the baroclinic term. Such an effect is widely considered to be 
instrumental 
in breaking the Taylor--Proudman balance in the solar
case (e.g., Rempel \cite{R05}; Miesch et al.\ \cite{MBT06}). 
Local simulations can be 
used to determine the latitudinal heat
flux but by virtue of periodic boundaries, no information about the
latitudinal profile of entropy can be
extracted from a single simulation.
Earlier local studies suggest that in the presence of
rotation the latitudinal heat flux is directed towards the poles
(e.g.\ R\"udiger et al.\ \cite{R05b}) and mean-field models in
spherical geometry indicate that such a flux leads to warm poles and
a cooler equator (e.g.\ Brandenburg et al.\ \cite{BMT92}), 
thus alleviating the Taylor--Proudman balance.
Computing the turbulent heat fluxes in spherical geometry in order
to compare with earlier results is one of the principal aims of the 
present study. Of particular importance is the sign and magnitude 
of the latitudinal heat flux.

It is possible that the use of Cartesian geometry and periodic
boundaries give rise to artefacts which are not present in fully
spherical geometry. In the present paper we undertake the computation
of Reynolds stress and turbulent heat transport from simulations in
spherical geometry as functions of rotation, and compare them with
Cartesian simulations of the same system located at different
latitudes. One of the most important goals of the paper is to find
out whether the present results in Cartesian geometry compare with
early similar studies and to test if these results are still valid
when spherical geometry is used. 
Earlier studies comparing spherical and Cartesian models used limited
two-dimensional geometry in the spherical case Hupfer et. al\
(\cite{HKS06}) whereas we perform all our simulations in three
dimensions. Furthermore, Robinson \& Chan (\cite{RC01}) used spherical
wedges to compute the rotation profiles and turbulent fluxes using two
representative runs. Here we explore a significantly larger portion of
parameter space. As a side result we also obtain angular velocity
profiles as a function of rotation from our spherical simulations
which, however, are dominated by the Taylor-Proudman balance in the
regime most relevant to the Sun. Thus we fail in reproducing the solar
rotation profile which is a common problem that can currently be
overcome only by introducing some additional poorly constrained terms,
e.g.\ a latitudinal entropy gradient, by hand rather than
self-consistently (e.g.\ Miesch et al.\ 2006).  Another important use
for the results will be the more ambitious future runs where
subgrid-scale models of the turbulent effects can be used to overcome
the Taylor--Proudman balance.

\section{Model}

Our spherical model is similar to that used by K\"apyl\"a et al.\
(\cite{KKBMT10}) but without magnetic fields. We model a segment of 
a star, i.e.\ a ``wedge'', in spherical polar
coordinates where $(r,\theta,\phi)$ denote the radius, colatitude, and
longitude. The radial, latitudinal, and longitudinal extents of the
computational domain are given by $0.65R \leq r \leq R$,
$\theta_0 \leq \theta \leq 180\degr-\theta_0$, and
$0 \leq \phi \leq \phi_0$,
respectively, where $R$ is the radius of the star.
In all of our runs we take $\theta_0=15\degr$ and
$\phi_0=90\degr$.
We study the dependence of the results on domain size in 
Appendix~\ref{depsize}.
In our Cartesian runs, the coordinates $(x, y, z)$ correspond to
radius, latitude and longitude of a box located at a colatitude
$\theta$.
Our domain spans from 
$0.65R \leq x \leq R$, $-0.35R \leq y \leq 0.35R$ and   
 $-0.35R \leq z \leq 0.35R$, i.e., the extension of the horizontal
directions is twice the vertical one, as has been used in
previous Cartesian simulations (e.g.\ K\"apyl\"a et al.\
\cite{KKT04}).

In both geometries, we solve the following equations
of compressible hydrodynamics,
\begin{equation}
\frac{D \ln \rho}{Dt} = -\bm\nabla\cdot\bm{u},
\end{equation}
\begin{equation}
\frac{D\bm{u}}{Dt} = \bm{g} -2\bm\Omega\times\bm{u}+\frac{1}{\rho}
\left(\bm\nabla \cdot 2\nu\rho\bm{\mathsf{S}}-\bm\nabla p\right),
\end{equation}
\begin{equation}
\frac{D s}{Dt} = \frac{1}{\rho T}\left(\bm\nabla \cdot K \bm\nabla T
+ 2\nu \bm{\mathsf{S}}^2 -\Gamma_{\rm cool}\right),
\label{equ:ss}
\end{equation}
where $D/Dt = \pd/\pd t + \bm{u} \cdot \bm\nabla$ is the advective time
derivative, 
$\nu$ is the kinematic viscosity, $K$ is the radiative heat conductivity, 
and $\bm{g}$ is the gravitational
acceleration given by
\begin{equation}
\bm{g}=-\frac{GM}{r^2}\hat{\bm{r}},
\end{equation}
where $G$ is the gravitational constant, $M$ is the mass of the star, and
$\hat{\bm{r}}$ is the unit vector in the radial direction.
Note that in the Cartesian case $x$ corresponds to the $r$
direction so that all radial profiles in spherical
coordinates directly apply to the Cartesian model.
We omit the centrifugal force in our models.
This is connected with the fact that the Rayleigh number is much less
than in the Sun, which is unavoidable and constrained by the numerical
resolution available.
This implies that the Mach number is larger than in the Sun.
Nevertheless, it is essential to have realistic Coriolis numbers. i.e.\
the Coriolis force has to be larger by the same amount that the turbulent
velocity is larger, but without significantly altering the hydrostatic
balance that is determined by gravity and centrifugal forces.

The fluid obeys the ideal gas law with $p=(\gamma-1)\rho e$, where
$\gamma=c_{\rm P}/c_{\rm V}=5/3$ is the ratio of specific heats in
constant pressure and volume, respectively, and $e=c_{\rm V} T$ is the
internal energy. The rate of strain tensor $\bm{\mathsf{S}}$ is given
by
\begin{equation}
\mathsf{S}_{ij}=\onehalf(u_{i;j}+u_{j;i})-\onethird \delta_{ij}\bm\nabla\cdot\bm{u},
\end{equation}
where the semicolons denote covariant differentiation (see Mitra et
al.\ \cite{MTBM09} for details).

The computational domain is divided into three parts: a lower
convectively stable layer at the base, convectively unstable layer and
a cooling layer at the top mimicking the effects of radiative losses
at the stellar surface.
The radial positions
$(r_1,r_2,r_3,r_4)=(x_1,x_2,x_3,x_4)=(0.65,0.7,0.98,1)R$ give the
locations of the 
bottom of the domain, bottom and top of the convectively unstable
layer, and the top of the domain, respectively.
The last term on the rhs of Eq.~(\ref{equ:ss}) describes cooling in
the surface layer given by
\begin{equation}
\Gamma_{\rm cool} = \Gamma_0 f(r) \left(\frac{\cst-c_{\rm s0}^2}{c_{\rm s0}^2}\right),
\end{equation}
where $f(r)$ is a profile function equal to unity in $r>r_3$ and
smoothly connecting to zero below, and $\Gamma_0$ is a cooling luminosity
chosen so that the sound speed in the uppermost layer relaxes
toward $c_{\rm s0}^2=\cst(r=r_4)$.

\subsection{Initial and boundary conditions}
\label{sec:initcond}
For the thermal stratification we adopt a simple setup that can be 
described analytically rather than
adopting profiles from a solar or
stellar structure model as in, e.g., Brun et al.\ (\cite{BMT04}).
We use a piecewise polytropic setup which
divides the domain into three layers. The hydrostatic temperature
gradient is given by
\begin{equation}
\frac{\pd T}{\pd r} = \frac{-g}{c_{\rm V}(\gamma-1)(n+1)},
\end{equation}
where $n=n(r)$ is the radially varying polytropic index. This gives
the logarithmic temperature gradient $\nabla$ (not to be confused with
the operator $\bm\nabla$) as
\begin{equation}
\nabla=\pd \ln T/\pd \ln p = (n+1)^{-1}.
\end{equation}
The stratification is unstable if $\nabla-\nabla_{\rm ad}>0$ where
$\nabla_{\rm ad}=1-1/\gamma$, corresponding to $n<1.5$. We choose
$n=6$ for the lower overshoot layer, whereas $n=1$ is used in the
convectively unstable layer.
A polytropic setup with $n=1$ is commonly used in convection studies
(e.g.\ Hurlburt et al.\ \cite{HTM84}).
This implies that about 80 per cent of the energy is transported
by radiation (cf.\ Brandenburg et al.\ \cite{BCNS05}), regardless
of the vigor of convection and the value of the Reynolds number.

\begin{figure}[t]
\centering
\includegraphics[width=\columnwidth]{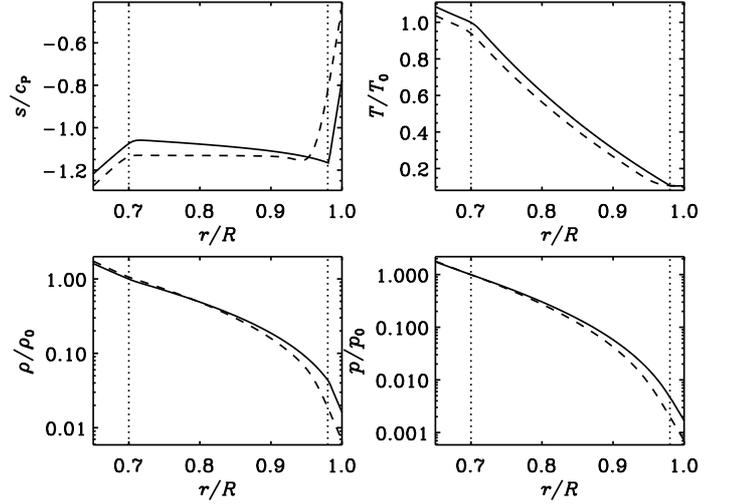}
\caption{Radial profiles of entropy, temperature, density, and
  pressure in the initial state (solid lines) and the in the saturated
  state (dashed) of Run~B0. 
  Reference values $T_0$ and $p_0$ are taken from the bottom of the
  convectively unstable layer in the initial state.
  The dotted vertical lines at $r_2=0.7R$
  and $r_3=0.98R$ denote the bottom and top of the convectively
  unstable layer, respectively.}
\label{fig:pther}
\end{figure}

Density stratification is obtained by
requiring hydrostatic equilibrium. The thermal conductivity is
obtained by requiring a constant luminosity $L_0$ throughout the domain
via
\begin{equation}
K=\frac{L_0}{4\pi r^2 \pd T/\pd r}.
\end{equation}
In order to expedite the initial transient due to thermal relaxation,
the thermal variables have a shallower profile, corresponding to
$\rho\propto T^{1.4}$, in the convection zone and $n=1$ is only used
for the thermal conductivity.
This gives approximately the right entropy jump that corresponds
to the required flux (cf.\ Brandenburg et al.\ \cite{BCNS05}).
In \Fig{fig:pther} we show the initial and final stratifications 
of specific entropy,
temperature, density, and pressure for a particular run.

In the spherical models the radial and latitudinal boundaries 
are taken to be impenetrable and stress free, according to
\begin{eqnarray}
\lefteqn{u_r=0,\quad \frac{\pd u_\theta}{\pd r}=\frac{u_\theta}{r},\quad \frac{\pd
u_\phi}{\pd r}=\frac{u_\phi}{r} \quad (r=r_1, r_4),}\\
\lefteqn{\frac{\pd u_r}{\pd \theta}=u_\theta=0,\quad \frac{\pd u_\phi}{\pd
\theta}=u_\phi \cot \theta \quad (\theta=\theta_0,\pi-\theta_0).}
\end{eqnarray}
On the latitudinal boundaries we assume that the thermodynamic
quantities have zero first derivative, thus suppressing heat fluxes
through the boundary.

In Cartesian coordinates we use periodic boundary conditions in
the horizontal directions ($y$ and $z$), and stress free conditions in
the $x$ direction, i.e.,
\begin{equation}
\lefteqn{u_x =  \frac{\pd u_y}{\pd x} = \frac{\pd
u_z}{\pd x}= 0 \quad (x=x_1, x_4).}
\end{equation}

The simulations were performed using the {\sc Pencil Code}%
\footnote{\texttt{http://pencil-code.googlecode.com/}},
which uses sixth-order explicit finite differences in space and a third-order
accurate time stepping method (see Mitra et al.\ \cite{MTBM09} for 
further information regarding the adaptation of the {\sc Pencil Code}
to spherical coordinates).

\begin{table*}[t!]
\centering
\caption[]{Summary of the spherical runs.}
% The runs are found in the folder:
% \texttt{pencil-code/petri/convection/spherical/turbtra}
      \label{tab:runs}
      \vspace{-0.5cm}
     $$
         \begin{array}{p{0.05\linewidth}cccccccccccc}
           \hline
           \noalign{\smallskip}
Run & $grid$ & \Ra  & \Pra & \mathcal{L} & \Ma & \Rey & \Co &
\tilde{E}_{\rm ther} & \tilde{E}_{\rm kin} &
E_{\rm mer}/E_{\rm kin} & E_{\rm rot}/E_{\rm kin} & \Delta\Omega/\Omega_{\rm eq}  \\ \hline 
A0 & 128\times256\times128 & 3.1\cdot10^6 & 1.0 & 1.4\cdot10^{-4} & 0.023 &
38 & 0.00 & 0.116 & 7.7\cdot10^{-5} & 0.045 & 0.004 & -     \\ % 128x256x128a
           \hline
A1 & 128\times256\times128 & 3.1\cdot10^6 & 1.0 & 1.4\cdot10^{-4} & 0.022 &
36 & 0.13 & 0.114 & 6.9\cdot10^{-5} & 0.016 & 0.022 & -0.15 \\ % 128x256x128a1
A2 & 128\times256\times128 & 3.1\cdot10^6 & 1.0 & 1.4\cdot10^{-4} & 0.022 &
36 & 0.25 & 0.114 & 7.2\cdot10^{-5} & 0.015 & 0.073 & -0.31 \\ % 128x256x128a2
A3 & 128\times256\times128 & 3.1\cdot10^6 & 1.0 & 1.4\cdot10^{-4} & 0.022 &
37 & 0.50 & 0.113 & 1.2\cdot10^{-4} & 0.010 & 0.438 & -1.03 \\ % 128x256x128a3
A4 & 128\times256\times128 & 3.1\cdot10^6 & 1.0 & 1.4\cdot10^{-4} & 0.029 &
48 & 0.94 & 0.112 & 1.1\cdot10^{-3} & 0.016 & 0.927 & -1.74 \\ % 128x256x128a4
A5 & 128\times256\times128 & 3.1\cdot10^6 & 1.0 & 1.4\cdot10^{-4} & 0.022 &
36 & 2.56 & 0.111 & 9.9\cdot10^{-4} & 0.002 & 0.949 & -0.37 \\ % 128x256x128a55
A6 & 128\times256\times128 & 3.1\cdot10^6 & 1.0 & 1.4\cdot10^{-4} & 0.018 &
30 & 6.09 & 0.114 & 2.3\cdot10^{-4} & 0.000 & 0.824 & +0.20 \\ % 128x256x128a6
           \hline
B0 & 128\times512\times256 & 8.6\cdot10^6 & 1.0 & 1.4\cdot10^{-4} & 0.020 &
54 & 0.00 & 0.113 & 5.8\cdot10^{-5} & 0.036 & 0.009 & -     \\ % 128x512x256a
           \hline
B1 & 128\times512\times256 & 8.6\cdot10^6 & 1.0 & 1.4\cdot10^{-4} & 0.020 &
57 & 1.34 & 0.112 & 6.5\cdot10^{-4} & 0.009 & 0.927 & -1.10 \\ % 128x512x256a3
B2 & 128\times512\times256 & 8.6\cdot10^6 & 1.0 & 1.4\cdot10^{-4} & 0.018 &
50 & 3.06 & 0.113 & 1.2\cdot10^{-4} & 0.001 & 0.689 & +0.12 \\ % 128x512x256a2
B3 & 128\times512\times256 & 8.6\cdot10^6 & 1.0 & 1.4\cdot10^{-4} & 0.016 &
44 & 6.93 & 0.113 & 1.8\cdot10^{-4} & 0.000 & 0.833 & +0.20 \\ % 128x512x256a1
           \hline
C1 & 128\times256\times128 & 1.7\cdot10^7 & 6.7 & 7.5\cdot10^{-5} & 0.008 &
12 & 7.58 & 0.114 & 3.8\cdot10^{-5} & 0.000 & 0.817 & +0.12 \\ % 128x256x128c6
           \hline
D1 & 256\times512\times256 & 6.0\cdot10^7 & 1.0 & 3.1\cdot10^{-5} & 0.012 &
90 & 5.07 & 0.113 & 7.9\cdot10^{-5} & 0.000 & 0.796 & +0.20 \\ % r256x512x256a1
D2 & 256\times512\times256 & 6.0\cdot10^7 & 1.0 & 3.1\cdot10^{-5} & 0.012 &
89 & 7.68 & 0.113 & 1.5\cdot10^{-4} & 0.000 & 0.895 & +0.20 \\ % r256x512x256a2
           \hline
         \end{array}
     $$
\tablefoot{
Here $\Ma=\urms/\sqrt{GM/R}$,
$\Delta \Omega = \Omega_{\rm eq}-\Omega_{\rm pole}$, 
where $\Omega_{\rm eq}=\mean{\Omega}(r_4,\theta=90\degr)$ and 
$\Omega_{\rm pole}=\mean{\Omega}(r_4,\theta=\theta_0)$.
$\tilde{E}_{\rm ther}=\brac{\rho e}$ and 
$\tilde{E}_{\rm kin}=\brac{\onehalf \rho \bm{u}^2}$ are the volume averaged
thermal and total kinetic energies, respectively, in units of $GM\rho_0/R$.
$E_{\rm mer}=\onehalf\brac{\rho(\mean{u}_\theta^2+\mean{u}_\phi^2)}$ 
and $E_{\rm rot}=\onehalf\brac{\rho\mean{u}_\phi^2}$ are the kinetic
energies of the meridional circulation and differential rotation.
}
\end{table*}

\subsection{Nondimensional quantities}
Dimensionless quantities are obtained by setting
\begin{eqnarray}
R = GM = \rho_0 = c_{\rm P} = 1\;,
\end{eqnarray}
where $\rho_0$ is the density at $r_2$,
The units of length, velocity, density, and entropy are then given by
\begin{eqnarray}
\lefteqn{ [x] = R\;,\;\; [u]=\sqrt{GM/R}\;,\;\;[\rho]=\rho_0\;,\;\;[s]=c_{\rm P}\;.}
\end{eqnarray}
The Cartesian simulations have been arranged so that the thickness
of the layers is the same, $\vec{g}=-(GM/x^2)\hat{\bm{x}}$, and $R$,
which is still our unit length, has no longer the meaning of a radius.
The simulations are governed by the Prandtl,
Reynolds, Coriolis, and Rayleigh numbers, defined by
\begin{eqnarray}
\Pra&\!=\!&\frac{\nu}{\chi_0}\;,\;\;\Rey=\frac{\urms}{\nu \kef}\;,\;\;
\Co=\frac{2\,\Omega_0}{\urms \kef}\;,\\
\Ra&\!=\!&\frac{GM(\Delta r)^4}{\nu \chi_0 R^2} \bigg(-\frac{1}{c_{\rm P}}\frac{{\rm d}s}{{\rm d}r} \bigg)_{r_{\rm m}}\;,
\label{equ:Co}
\end{eqnarray}
where $\chi_0 = K/(\rho_{\rm m} c_{\rm P})$ is the thermal
diffusivity, $\kef=2\pi/\Delta r$ is an estimate of the wavenumber of
the energy-carrying eddies, $\Delta r=r_3-r_2$ is the thickness of
the unstable layer, $\rho_{\rm m}$ is the density in the
middle of the unstable layer at \ $r_{\rm m}=(r_3+r_2)/2$,
and $\urms=\sqrt{\threehalfs\brac{u_r^2+u_\theta^2}}$ is
the rms velocity, where the angular brackets denote volume
averaging. In our definition of $\urms$ we omit the contribution from
the $\phi$-component of velocity, because it is dominated by the large-scale
differential rotation that develops when rotation is included.
The entropy gradient, measured at $r_{\rm m}$ in the initial
non-convecting state, is given by
\begin{eqnarray}
\bigg(-\frac{1}{c_{\rm P}}\frac{{\rm d}s}{{\rm d}r}\bigg)_{r_{\rm m}} = \frac{\nabla_{\rm m}-\nabla_{\rm ad}}{H_{\rm P}}\;,
\end{eqnarray}
where $\nabla_{\rm m} = (\pd \ln T/\pd \ln p)_{r_{\rm m}}$, 
and $H_{\rm P}$ is the pressure scale height at $r_{\rm m}$.

\begin{figure}[t]
\centering
\includegraphics[width=\columnwidth]{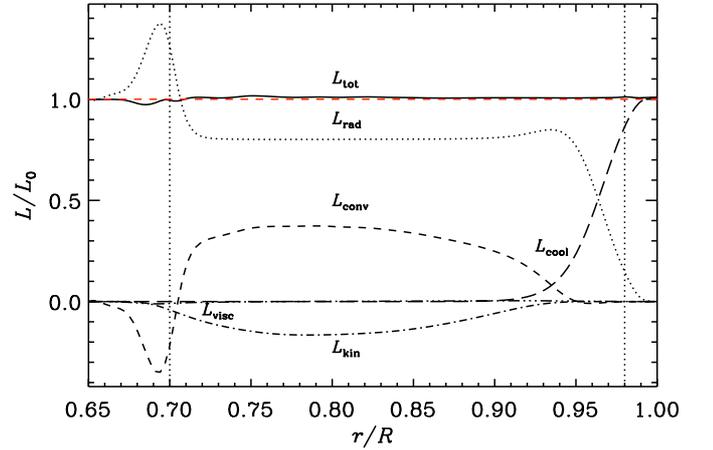}
\caption{Radiative (dotted line), enthalpy (dashed), kinetic energy
  (dash-dotted), cooling (long dashed), and viscous
  (triple-dot-dashed) luminosities as functions of radius from
  Run~A0. The solid line shows the sum of all fluxes, and the red 
  dashed line the luminosity $L_0$ fed into the domain through
  the lower boundary. The vertical dotted lines at $r=0.7R$ and
  $r=0.98R$ denote respectively the bottom and top of the convectively
  unstable layer in the initial state.}\label{fig:pflux_A0}
\end{figure}

\begin{figure*}[t!]
\centering
\includegraphics[width=\textwidth]{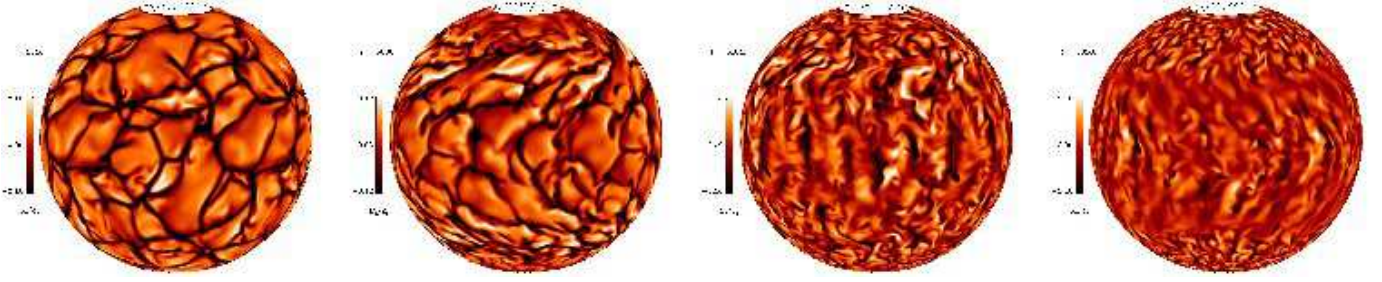}
\caption{Radial velocity $u_r$ at a small distance ($r=0.9R$) below the surface
  from Runs~B0--B3.
  The scales give $u_r$ in units of the local sound speed.
  For visualization purposes, the domain is duplicated fourfold in the
  longitudinal direction.
  See also \texttt{http://www.helsinki.fi/\ensuremath{\sim}kapyla/movies.html}}
\label{fig:c128x512x256a5_1000}
\end{figure*}

The energy that is deposited into the domain at the base is
controlled by the luminosity parameter
\begin{equation}
\mathcal{L} = \frac{L_0}{\rho_0 (GM)^{3/2} R^{1/2}},
\end{equation}
where $L_0=4\pi r_1^2 F_{\rm b}$ is the constant luminosity, and 
$F_{\rm b}=-(K \pd T/\pd  r)|_{r=r_1}$ is the energy flux imposed at 
the lower boundary.
Furthermore, the stratification is determined by the pressure scale
height at the surface
\begin{eqnarray}
\xi = \frac{(\gamma-1) c_{\rm V}T_4}{GM/R},
\end{eqnarray}
where $T_4=T(r=r_4)$.
Similar parameter definitions were used by Dobler et al.\
(\cite{DSB06}).
We use $\xi=0.020$, which results in a density contrast of $10^2$
across the domain.

\section{Results}

Our main goal is to extract the turbulent fluxes of angular momentum
and heat as functions of rotation from our simulations. In order to
achieve this we use a moderately turbulent model
and vary the rotation rate,
quantified by the Coriolis number, from zero to roughly six in Set~A
(see Table~\ref{tab:runs}). We also perform a subset of these
simulations at higher resolution in Set~B and a three runs (C1, 
D1, and D2) with a lower Mach number. 
The runs in Set~A were initialized
from scratch, whereas in Set~B a nonrotating simulation B0 was run until
it was thermally relaxed. 
The runs with rotation (B1--B3) were
then started from this snapshot and computations carried out until a
new saturated state was reached. 
The runs D1 and D2 were remeshed from a non-rotating, thermally 
relaxed model at a lower resolution.
In \Fig{fig:pther} we compare the
initial and final stratification of specific entropy, temperature,
density, and pressure for Run~B0.

As noted in Sect.~\ref{sec:initcond}, our polytropic setup leads to a
system where radiative diffusion transports 80 per cent of the total 
energy. We show the flux balance in the statistically saturated state 
from Run~A0 in Fig.~\ref{fig:pflux_A0}, where the different 
contributions are given in terms of luminosities $L_i=4\pi r^2 F_i$,
and where
\begin{eqnarray}
F_{\rm rad} &=& - K\frac{\pd T}{\pd r}, \\
F_{\rm conv} &=& - c_{\rm P}\mean{\rho} \mean{u_r' T'}, \\
F_{\rm kin} &=& \onehalf \mean{\rho} \mean{u^2 u_r}, \\
F_{\rm visc} &=& -2\nu \mean{\rho}\ \mean{u_i\mathsf{S}_{ir}}, \\
F_{\rm cool} &=& \int \Gamma_{\rm cool}\ dr.
\end{eqnarray}
Here we consider averages over $\phi$ and $\theta$.
We find that in the non-rotating case the convective flux accounts for
roughly 30 per cent of the total luminosity and the (inward) kinetic
energy flux is between 10 and 15 per cent. When rotation is increased,
both the convective and kinetic fluxes decrease. The viscous flux is
always negligible. The cooling flux transports the total luminosity
near the surface.

Visualizations of $u_r$ at a small distance below the surface are shown in
\Fig{fig:c128x512x256a5_1000} for Runs~B0--B3. 
The convective velocities $\vec{u}'$ can be decomposed in terms
of poloidal ($\vec{u}'_{\rm P}$) and toroidal ($\vec{u}'_{\rm T}$) parts 
following Lavely \& Ritzwoller (\cite{LR92})
\begin{equation}
\label{eq:sph1}
\vec{u}'_{\rm P}={\rm Real} \sum_{l,m} \left\lbrace u_{\rm P}^{lm}(r)Y_l^m\ru + 
v_{\rm P}^{lm}(r)\ve{\nabla}Y_l^m\right\rbrace, %\\
\end{equation} 
\begin{equation}
\label{eq:sph2}
\vec{u}'_{\rm T}={\rm Real} \sum_{l,m}\left\lbrace w_{\rm T}^{lm}(r)\ru \times\ve{\nabla}Y_l^m \right\rbrace,
\end{equation} 
where $Y_l^m(\theta, \phi)$ are spherical harmonics of degree $l$ and order $m$.
The geometry and amplitude of 
the poloidal velocity are completely defined by $l$, $m$, and $u_{\rm P}^{lm}$ 
since, assuming approximate mass conservation, 
$v_{\rm P}^{lm}$ and $u_{\rm P}^{lm}$ are related as
\begin{equation}
v_{\rm P}^{lm}(r) = \frac{\partial_r(r^2\rho u_{\rm P}^{lm}(r))}{\rho r l(l+1)}.
\end{equation}
The poloidal flow has characteristics of B\'enard convection cells with 
upwellings at the centres of cells and downdraughts on the peripheries.
The toroidal flows are characterised by their amplitude and geometry given by 
$w_{\rm T}^{lm}$, $l$, and $m$ respectively. In contrast to poloidal flows, 
their nature resembles that of rotation, jets or horizontal vortices.
In \Fig{fig:c128x512x256a5_1000}, we observe that so called {\em banana cells} become
prominent in the radial velocity with an increase in the Coriolis number. 
Such structures are poloidal
flows given by spherical harmonic $Y_l^m(\theta, \phi)$. For Run~B3 in 
\Fig{fig:c128x512x256a5_1000}, we find maximum power at $m = 16$. 
Note that the reality of the banana cells in the Sun is hotly
debated. Even though significant power is found at wavenumbers corresponding to giant cells
in the surface velocity spectra of the Sun, 
no distinct peak has been found at those 
wavenumbers (Chou et al.\ \cite{Cea91}; Hathaway et al.\ \cite{Hea00}). 
Global helioseismology 
caps the maximum radial
velocity of the banana cells at $50$~m~s$^{-1}$ (Chatterjee \& Antia
\cite{CA09}).
We study the importance of the banana cells to the Reynolds stresses 
in more detail in Sects.~\ref{sec:Rey} and \ref{sec:ban}.

\begin{figure}[t]
\centering
\includegraphics[width=0.95\columnwidth]{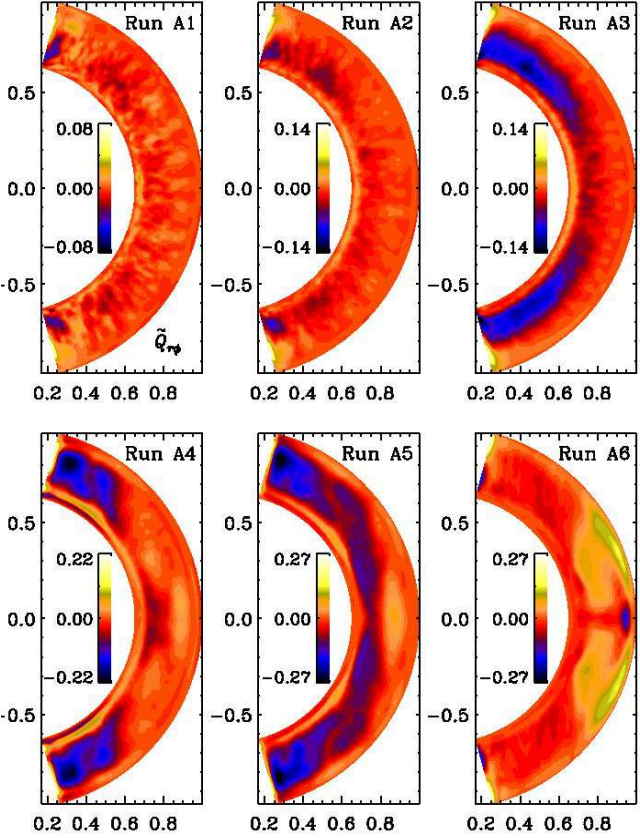}
\caption{Vertical Reynolds stress, $\tilde{Q}_{r\phi}$, from Set~A.
}\label{fig:pRxzA}
\end{figure}

\subsection{Reynolds stress}
\label{sec:Rey}
The angular momentum balance of a star is governed by the conservation
law (R\"udiger \cite{R89})
\begin{equation}
\frac{\pd}{\pd t} (\mean{\rho}\varpi^2 \mean{\Omega}) =-\bm\nabla\cdot
\left[\mean{\rho}\varpi\left(\varpi \mean{\Omega} \meanv{u}_{\rm mer}
+ \mean{u_\phi' \bm{u}'}\right)\right],
\end{equation}
where $\varpi=r \sin \theta$ is the lever arm and $\meanv{u}_{\rm
  mer}=(\mean{u}_r,\mean{u}_\theta)$ is the meridional
circulation. The latter term on the rhs describes the effects of the
Reynolds stress components $\qrp$ and $\qtp$, which describe radial and
latitudinal fluxes of angular momentum, respectively. The stress is
often parameterised by turbulent transport coefficients that couple
small-scale correlations with large-scale quantities, i.e.\
\begin{equation}
\qij=\Lambda_{ijk}\mean{\Omega}_k-\mathcal{N}_{ijkl}\frac{\pd\mean{u}_{k}}{\pd x_l},
\end{equation}
where $\Lambda_{ijk}$ describes the nondiffusive contribution
($\Lambda$-effect) and $\mathcal{N}_{ijkl}$ the diffusive part
(turbulent viscosity), cf.\ R\"udiger (\cite{R89}). However,
disentangling the two contributions is not possible, see e.g.,
Snellman et al.\ (\cite{SKKL09}) and K\"apyl\"a et al.\
(\cite{KBKSN10}). We postpone a detailed study of the turbulent transport
coefficients to a future study and concentrate on comparing the total
stress with simulations in Cartesian geometry.

\begin{figure}[t]
\centering
\includegraphics[width=0.95\columnwidth]{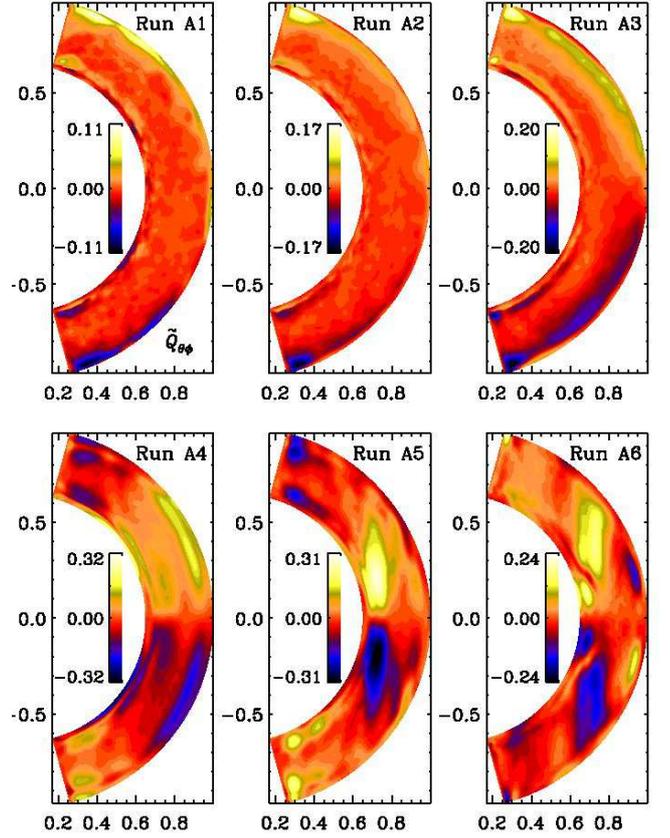}
\caption{Horizontal Reynolds stress, $\tilde{Q}_{\theta\phi}$, from Set~A.
}\label{fig:pRyzA}
\end{figure}

It is convenient to display the components of the Reynolds stress in
non-dimensional form (indicated by a tilde), and to define
\begin{equation}
\label{eq:qij}
\tilde{Q}_{ij}=\mean{u_i' u'_j}/\urms^2,
\end{equation}
where $\urms=\urms(r,\theta)$ is the meridional rms-velocity.
The averages are calculated over the azimuthal direction and time
also for $\urms$.
In the following, we refer to the three off-diagonal components,
$Q_{r\phi}$, $Q_{\theta\phi}$, and $Q_{r\theta}$,
as vertical, horizontal, and meridional components, respectively.
Representative results for the vertical stress component $\qrp$ are
shown in Fig.~\ref{fig:pRxzA}. We find that for slow rotation
(Run~A1), $\qrp$ is small and does not appear to show a clear trend in
latitude. In Run~A2 with $\Co\approx0.25$ the stress is more
consistently negative within the convectively unstable layer, showing
a symmetric profile with respect to the equator.
These two runs tend to show the largest signal near the latitudinal
boundaries which is most likely due to the boundary conditions
there. Similar distortions are also seen in the large-scale flows (see
Sect.~\ref{sec:LS}). In the intermediate rotation regime
(Runs~A3--A5), $\qrp$ is predominantly negative, although regions of
opposite sign start to appear near the equator. In Run~A6 the stress
is mostly positive.
Qualitatively similar results are obtained from the runs in Set~B,
Runs~C1, D1, and D2.
Therefore there is a sign change roughly at $\Co=2$. 
The results for most quantities from Runs~B2 and D1 with 
intermediate values of $\Co$ are similar to those of Runs~A5 and A6, 
respectively. Thus, we usually show results only from Runs~A4, A5, 
and A6 in order to demonstrate the qualitative change that occurs 
for many quantities in the range $\Co\approx1\ldots6$.
A similar phenomenon 
has been observed in
Cartesian simulations (K\"apyl\"a et al.\ \cite{KKT04}).
We note that the behaviour of $\qrp$ in the most rapidly rotating runs, 
namely a small negative region at the equator and a positive peak near 
the surface at somewhat higher latitudes was also reported by 
Robinson \& Chan (\cite{RC01}).

\begin{figure*}[t]
\centering
\includegraphics[width=.9\textwidth]{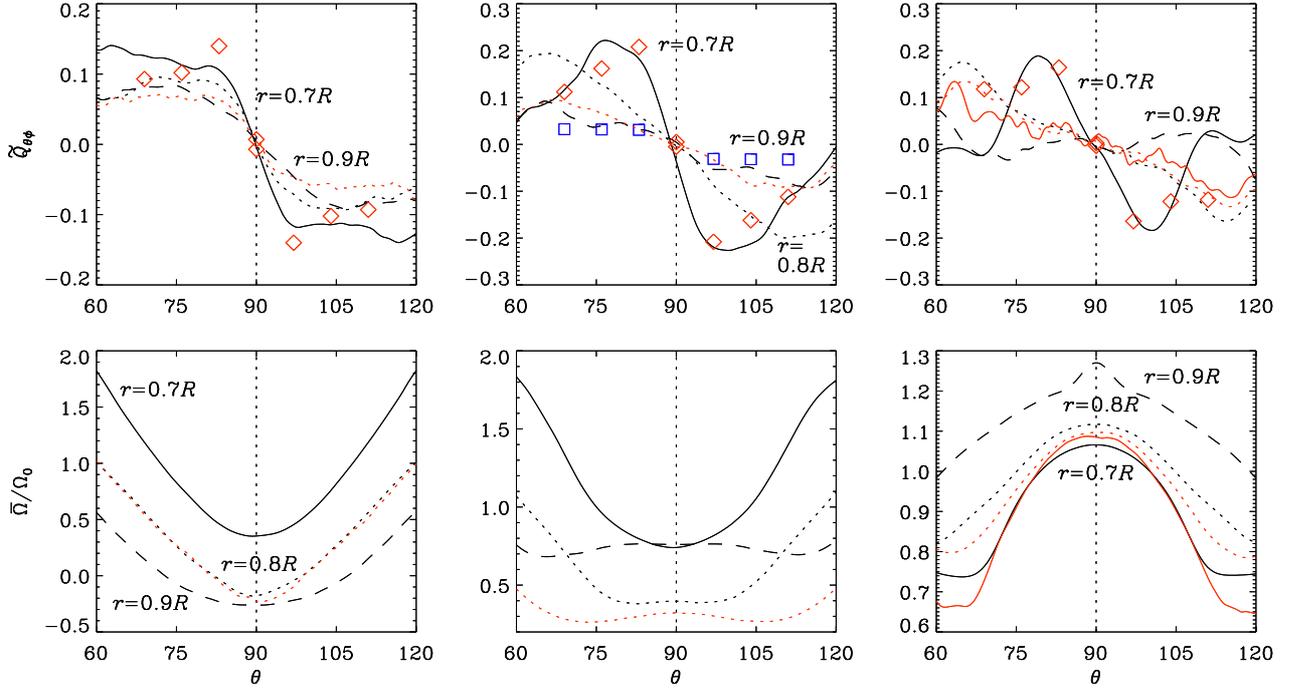}
\caption{Latitudinal profiles of $\tilde Q_{\theta\phi}$ and
  $\tilde\Omega$ for Runs~A4, A5, and A6 (from left to right) at three 
  different depths (solid 0.7$R$, dotted 0.8$R$, dashed 0.9$R$).
  The red dotted lines show data from corresponding Runs~B1, B2, and 
  B3 from $r=0.8R$.
  The solid red lines in the right panels show data from Run~D2 at
  $r=0.8R$.
  The
  open red diamonds in the top panels denote Cartesian Runs~cA1--cA4,
  cD1--cD4, and cE1--cE4, from left to right. The blue squares in
  the top-middle panel show the values of $Q_{yz}$ computed from 
  Fourier-filtered velocity fields
  from Runs~cD1--cD4.
  Note that only a part of the full latitudinal range is shown.
}\label{fig:pline_QhorOm_A5}
\end{figure*}

We find that the horizontal stress, $\tilde Q_{\theta \phi}$, is
almost always positive (negative) in 
the northern (southern) hemisphere for $\Co<1$, 
i.e. antisymmetric about the equator,
see Fig.~\ref{fig:pRyzA}. For intermediate rotation (Runs~A4 and A5) 
the stress is observed to change sign at high latitudes. 
In Fig.~\ref{fig:pline_QhorOm_A5}
we plot the latitudinal profiles of the horizontal stress
and the mean angular velocity at different depths for the Runs~A4--A6. 
It can be seen that near the bottom of the convection
zone, the profile of the stress becomes more and more concentrated
about the equator as the Coriolis number increases.
An especially abrupt
change can be observed for Run~A5 ($\Co\approx2$).
A similar peak also persists in
Runs~A6, B3, C1, and D2 with the largest Coriolis numbers.
Note, however, that the sign of the latitudinal differential rotation
changes as $\Co$ increases to six for Run~A6.
The results of Robinson \& Chan (\cite{RC01}) also show a peak 
of $\qtp$, occurring at a latitude range $10\degr\ldots 15\degr$, depending on
depth.

Using Eqs.~(\ref{eq:sph1})--(\ref{eq:sph2}), we can calculate the stress
$Q_{\theta\phi}=\sum_{l,l',m}Q_{\theta\phi}^{ll'm}$ by azimuthal averaging, with
\begin{equation}
\label{eq:qtpkkl}
\nonumber
Q_{\theta\phi}^{ll'm}=\frac{1}{2}v_{\rm P}^{lm}w_{\rm T}^{l'm}\left(\frac{1}{r^2}
\frac{\partial P_l^m}{\partial\theta}\frac{\partial P_{l'}^m}
{\partial\theta}-\frac{m^2}{\varpi^2}P_l^mP_{l'}^m\right),
\end{equation}
where $P_l^m(\theta)$ are the associated Legendre polynomials and $\varpi=r\sin\theta$. 
Note that $l$ and $l'$ denote the degrees of the poloidal and the
toroidal flow, respectively.
It is easy to see that the contribution
to the azimuthally averaged $Q_{\theta\phi}$ is always zero from 
cross-correlation between two poloidal velocity fields. Finite contributions to $Q_{\theta\phi}$ 
instead come from correlations between poloidal flow and toroidal flow having 
the same azimuthal degree $m$. 
We have used small-scale velocity fluctuations (i.e., $m\neq0$ modes) 
to calculate the Reynolds stresses in the numerical simulations 
according to Eq.~(\ref{eq:qij}). The finite correlation of the rotation and 
the meridional flow are not included in this discussion since both 
are characterised by $m=0$ and thus do not correspond to our definition of 
velocity fluctuations.

\begin{figure}
\centering{\includegraphics[width=0.5\textwidth]{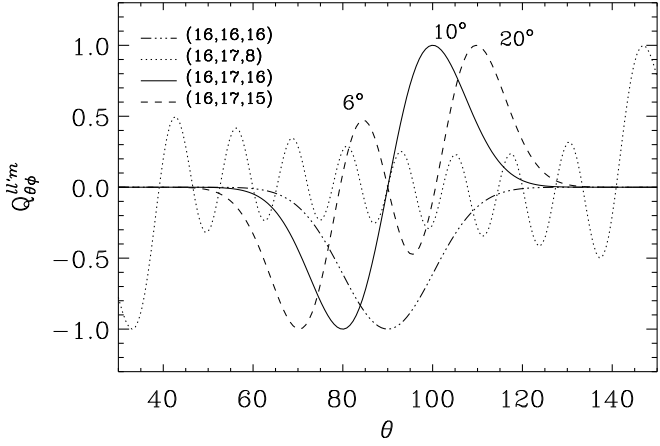}}
\caption{Angular part of $Q_{\theta\phi}^{ll'm}$
  normalized by the maximum value for four different cases
  characterized by triplets $(l, l', m)$ as indicated by the
  legend. The latitudes of the peaks for the triplets are indicated on
  the respective curves.}
\label{fig:sph}
\end{figure}

Recently, Bessolaz \& Brun\ (\cite{BB11}) have used wavelets and
autocorrelation techniques to unravel the structure of giant cells in
their 3-dimensional hydrodynamic convection simulations.  It is an
involved exercise to calculate the net stress by estimating the power
in each triplet $(l, l', m)$ by wavelet analysis. It is, however,
possible to look for certain combinations of Legendre polynomials that
can contribute to the peaks of $Q_{\theta\phi}$ near the equator as
obtained from numerical simulations in spherical geometry. A visual
inspection of the radial flows in Fig.~\ref{fig:c128x512x256a5_1000}
for Run~B2 shows four prominent banana cells within the domain which
extends from $0$ to $\pi/4$ in the azimuthal direction, which means
that the angular dependence is most likely $Y_{16}^{16}$. Hence we set
$l=16, l'=16, 17$ for the calculation of the stresses and vary $m$ in search
for a match between the peaks of $Q_{\theta\phi}$ from the runs A1--A6
and Eq.~(\ref{eq:qtpkkl}). We illustrate the angular part of
$Q_{\theta\phi}^{ll'm}$, for particular values of $l, l'$ and $m$ in
Fig.~\ref{fig:sph}. We can see from here that peaks in
$Q_{\theta\phi}^{16,17,15}$ (dashed line) appear at $\pm6^{\circ}$ as
well as at $\pm 20^{\circ}$ latitude, whereas peaks in
$Q_{\theta\phi}^{16,17,16}$ appear at $\pm 10^{\circ}$ latitude, and
the highest peaks in $Q_{\theta\phi}^{16, 17, 8}$ appear at $\pm
60^{\circ}$ latitude.  Comparing Fig.~\ref{fig:pRyzA} with
Fig.~\ref{fig:sph}, we see that at slow rotation (Runs~A1 and A2), a
major contribution to the stress may come from giant cells with an
angular dependence $Y_{16}^8$.  At higher $\rm Co$, the stress may
have contributions from banana cells with angular dependence
$Y_{16}^{16}$ (compare solid line in top right panel of
Fig.~\ref{fig:pline_QhorOm_A5} with solid line of Fig.~\ref{fig:sph}).
We shall return to the question regarding the contribution of banana
cells in the context of Cartesian runs in Sect.~\ref{sec:ban}.
However there also exists symmetric contribution to $Q_{\theta\phi}$
from components like $Q_{\theta\phi}^{16,16,16}$, but we do not see
any significant symmetric part in the horizontal stresses from the
numerical simulations. On this basis, zonal flows of the form 
$w_{\rm  T}^{ll}\ru\times\ve{\nabla}Y_l^l$ can be said to be negligible in
spherical convection simulations. These zonal flows correspond to a
row of horizontal vortices with their centres on the equator.

Finally, let us discuss the stress component $\qrt$.
It does not directly contribute to
angular momentum transport, but it can be important in generating
or modifying meridional circulation, and it has routinely been
considered also in earlier studies (e.g., Pulkkinen et al.\ \cite{Pea93};
Rieutord et al.\ \cite{RBMD94}; K\"apyl\"a et al.\ \cite{KKT04}).
Figure~\ref{fig:pRxyA} shows the
stress component $\qrt$ from Set~A. We find that for slow
rotation (Run~A1) the stress is quite weak and shows several sign
changes as a function of latitude. It is not clear whether this pattern
is real or an artefact of insufficient statistics. For intermediate
rotation (Runs~A2--A4), $\qrt$ shows an antisymmetric profile with
respect to the equator being positive in the northern hemisphere and
negative in the south, in accordance with earlier Cartesian results
(e.g.\ K\"apyl\"a et al.\ \cite{KKT04}). Although the theory for this
stress component is not as well developed as that of the other two
off-diagonal components, R\"udiger et al.\ (\cite{R05a}) state that
$\qrt$ should always be negative in the northern hemisphere, which is
at odds with our results. However, in our rapid rotation models
(Runs~A5--A6) the sign is found to change.

\subsection{Comparison with Cartesian simulations}

Before describing the Reynolds stress obtained from
our simulations in Cartesian coordinates, we note that the rms
velocities in the Cartesian runs are in general almost twice as large as in
the spherical ones with the same input parameters (compare, e.g.,
Run~A0 in Table~\ref{tab:runs} and Run~cA0 in
Table~\ref{tab:car_runs}). We argue in Sect.~\ref{sec:tuhe} that this
is the result of adopting a radial dependence of gravity in the
plane-parallel atmosphere.

\begin{figure}[t]
\centering
\includegraphics[width=\columnwidth]{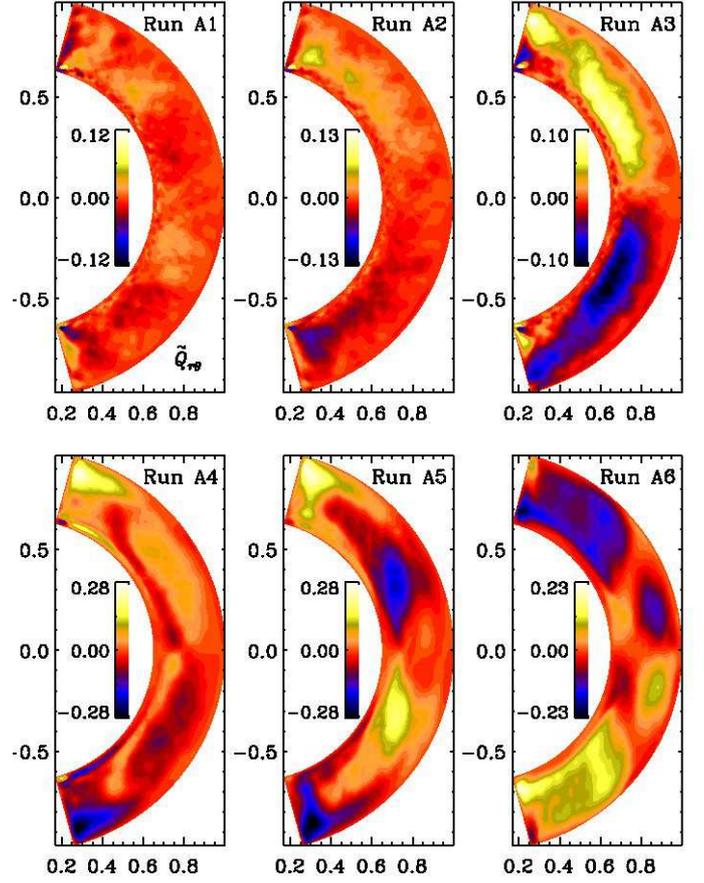}
\caption{Meridional Reynolds stress, $\tilde{Q}_{r\theta}$, from Set~A.
}\label{fig:pRxyA}
\end{figure}

\begin{table}[t]
\centering
\caption[]{Summary of the runs in Cartesian coordinates.} 
      \label{tab:car_runs}
      \vspace{-0.5cm}
     $$
         \begin{array}{p{0.05\linewidth}ccccccc}
           \hline
           \noalign{\smallskip}
Run & $Latitude$ &  \Rey & \Co & \Ma 
& \tilde E_{\rm k} & E_{\rm mer}/E_{\rm k}
& E_{\rm rot}/E_{\rm k}   \\ \hline   
cA0 & - & 63 & 0.00 &  0.038 &  1.7\cdot10^{-4} &  0.052 & 0.001
\\%c128_0d_0.0O
cF0 & - & 28 & 0.00 &  0.027 &  2.3\cdot10^{-4} &  0.021 & 0.001
\\%c128_Fcte_0.03
cF1 & - & 11 & 0.00 &  0.022 &  2.5\cdot10^{-4} &  0.002 & 0.000
\\%c128_Fcte_0.02
\hline
cA1 & 0^{\circ} & 64 & 0.85 & 0.039 & 2.9\cdot10^{-4} &  0.001 &  0.288
\\%c128_0d_0.6O
cA2 & 7^{\circ} & 65 &  0.84 & 0.039 & 2.0\cdot10^{-4} & 0.021 &  0.017
\\%c128_7d_0.6O
cA3 &14^{\circ} & 65 & 0.84 &  0.039 & 1.8\cdot10^{-4} &  0.014  &  0.007
\\%c128_14d_0.6O
cA4 & 21^{\circ} & 65 &  0.85 & 0.039 & 1.8\cdot10^{-4} & 0.012 & 0.008
\\%c128_21d_0.6O
\hline
cB1 & 0^{\circ} & 61 & 1.49 &  0.037 & 4.5\cdot10^{-4} &  0.000 &  0.623
\\%c128_0d_1.0O
cB2 & 7^{\circ} & 70 &  1.30 &  0.042 & 2.4\cdot10^{-4} &  0.023 & 0.012
\\%c128_7d_1.0O
cB3 &14^{\circ} & 68 & 1.33 & 0.041 & 2.0\cdot10^{-4} &  0.012 &  0.007
\\%c128_14d_1.0O
cB4 & 21^{\circ} & 68 & 1.34 &  0.041 & 1.9\cdot10^{-4} & 0.005 & 0.009
\\%c128_21d_1.0O
\hline
cC1 & 0^{\circ} & 60 & 2.14 &  0.036 & 2.8\cdot10^{-4} & 0.000 &  0.347
\\%c128_0d_1.4O
cC2 & 7^{\circ} & 76 & 1.68 &  0.046 & 2.5\cdot10^{-4} & 0.029 &  0.031
\\%c128_7d_1.4O
cC3 &14^{\circ} & 72 & 1.77 &  0.044 & 2.2\cdot10^{-4} &  0.013 &  0.011
\\%c128_14d_1.4O
cC4 &21^{\circ} & 72 &  1.78 &  0.043 & 2.1\cdot10^{-4} &  0.004 &  0.011
\\%c128_21d_1.4O
\hline
cD1 & 0^{\circ} & 69 & 2.38 & 0.042 & 7.5\cdot10^{-4} &  0.000 & 0.584
\\%c128_0d_1.8O
cD2 & 7^{\circ} & 78 & 2.09 & 0.047 &2.5\cdot10^{-4} &  0.029 &  0.018
\\%c128_7d_1.8O
cD3 & 14^{\circ} & 47 & 2.32 & 0.043 & 2.0\cdot10^{-4} & 0.009 &  0.013
\\ %c128_14d_1.8O
cD4 &21^{\circ} & 70 & 2.36 &  0.042 &2.1\cdot10^{-4} & 0.003 &  0.005
\\%c128_21d_1.8O
\hline
cE1 & 0^{\circ} & 50 &  3.66 &  0.045 &1.2\cdot10^{-3}  &  0.000 &  0.685
\\%c128_0d_3.0O
cE2 & 7^{\circ} & 36 &  4.00 &  0.041 & 1.6\cdot10^{-4}  &  0.025 &  0.009
\\%c128_7d_3.0O
cE3 & 14^{\circ} & 34  &  4.24 &  0.039 & 1.5\cdot10^{-4} &  0.005 &  0.005
\\ %c128_14d_3.0O
cE4 &21^{\circ} & 31 & 4.67 & 0.035 & 1.3\cdot10^{-4} & 0.001 &  0.008
\\%c128_21d_3.0O
\hline
\end{array}
     $$
\tablefoot{
  Here, we use a resolution of $64\times 128^2$ grid points.
  For the sets of Runs~cA--cD, $\Ra \approx 3.1 \cdot 10^6$, and for
  the set of Runs~cE, $\Ra \approx 1.4 \cdot 10^6$. Thermal energy in all 
  of the cases is $\tilde E_{\rm ther} \approx 0.117$. All quantities
  are computed using the same definitions and normalization factors as in
  Table \ref{tab:runs}. 
}
\end{table}

\begin{figure*}[t]
\centering
\includegraphics[width=.9\textwidth]{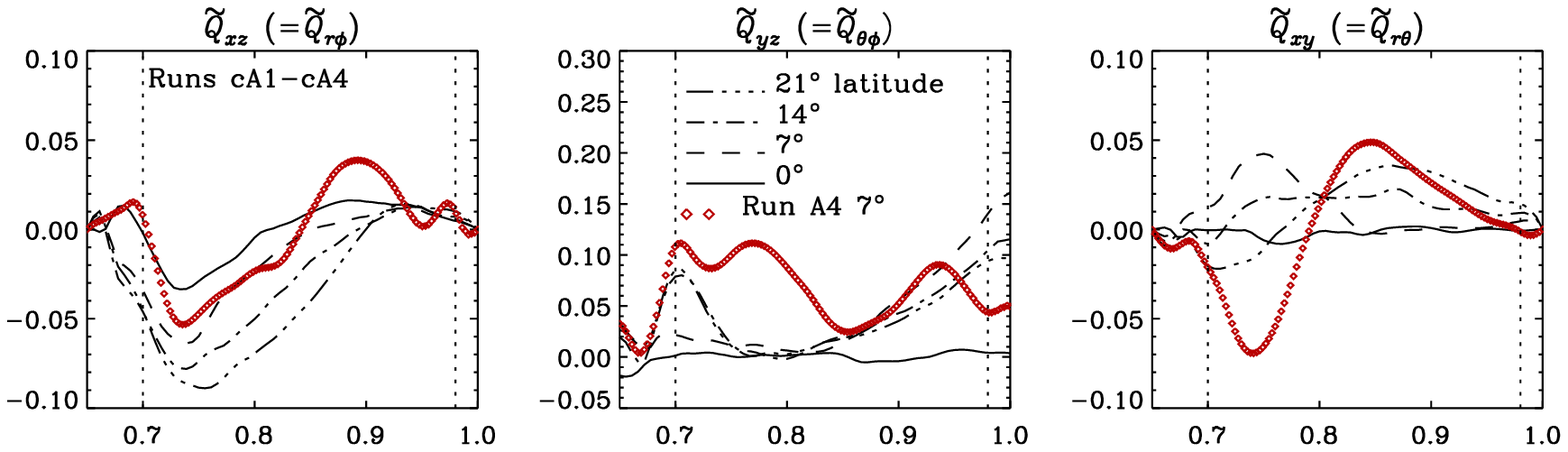}
\includegraphics[width=.9\textwidth]{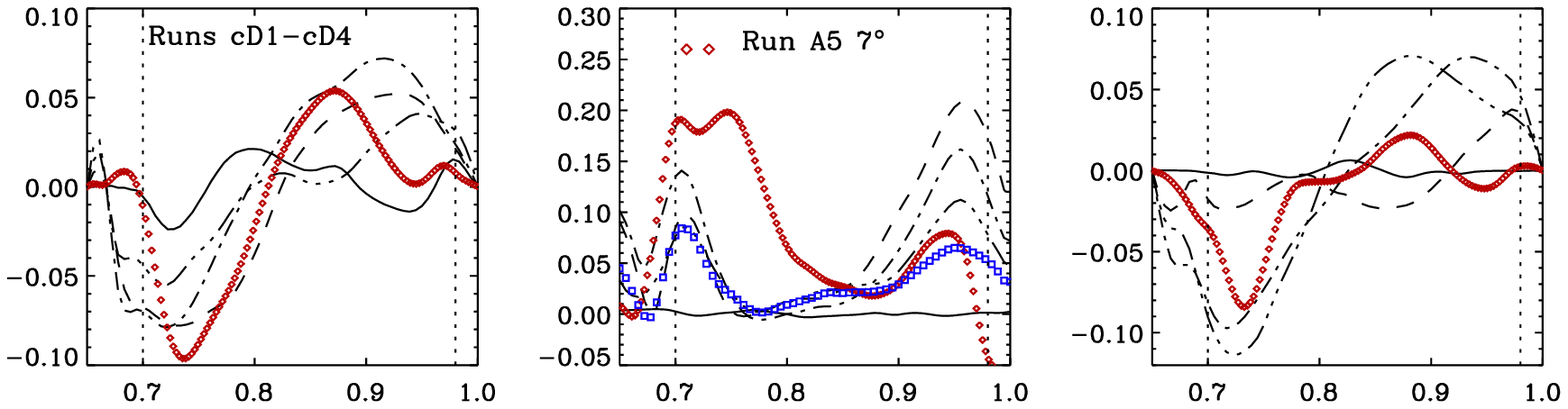}
\includegraphics[width=.9\textwidth]{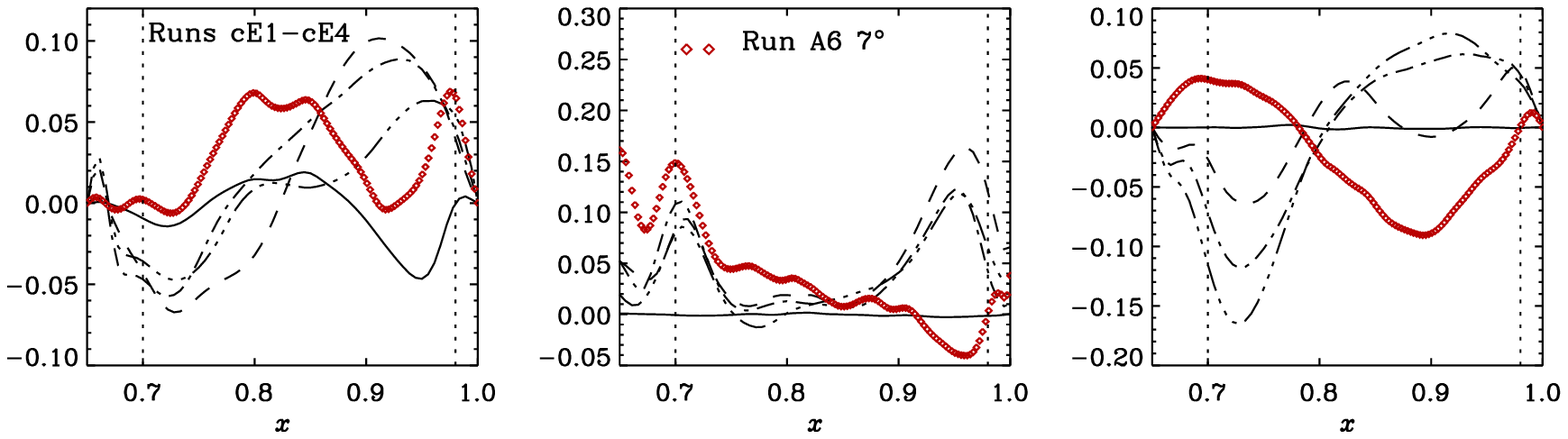}
\caption{From left to right: radial profiles of $\tilde Q_{xz}$, 
$\tilde Q_{yz}$, and $\tilde Q_{xy}$ from Cartesian Runs~cA1--cA4
(top panels), Runs~cD1--cD4 (middle panels), and Runs~cE1--cE4 
(bottom panels). 
The red diamonds correspond to the radial
profiles of the stresses in the spherical Runs~A4--A6. 
The blue squares in the middle panel show Fourier-filtered data
from Run~cD2.
}
\label{fig:cRss}
\end{figure*}

The radial profiles of the three off-diagonal components of the 
Reynolds stress in Cartesian
coordinates agree with previous studies (K\"apyl\"a et al.\ 
\cite{KKT04}; Hupfer et al.\ \cite{HKS05}) for
the range of latitudes and  Coriolis number explored here (compare
Fig.~\ref{fig:cRss} with bottom panel of Fig. 11 of K\"apyl\"a et al.\
\cite{KKT04} and Figs.~3 and 5 of Hupfer et al.\ \cite{HKS05}).        
For moderate rotation (Runs~cA1--cA4), the vertical
component $\tilde Q_{xz}$ (left panels of Fig.~\ref{fig:cRss}) is
negative in the bottom part of the convection zone and almost zero at
the top. The cases with $\Co\approx2.3$ (Runs~cD1--cD4) show negative
values at the bottom and positive values at the top of the convection
zone. For $\Co\approx4.0$ (Runs~cE1--cE4), the amplitude of the positive
part of the stress near the surface increases and the
negative part at the bottom decreases. We notice that the spatial
distribution of $\tilde Q_{xz}$, as well as its variation with the
Coriolis number, are in a fair agreement with the corresponding
spherical runs in the same range of $\Co$ (Runs~A3--A5). 
In the spherical Run~A6 with the highest Coriolis number of roughly six,
the stress is observed to become predominantly positive in the convection
zone. This is not seen in the Cartesian counterparts that reach
Coriolis numbers of roughly four (Runs~cE1--cE4), in which the
negative peak near the bottom still persists, although it has decreased
in magnitude.
The difference is possibly due to the lower Coriolis number in the 
Cartesian runs.
It is noteworthy that also the symmetry of this stress component with
respect to the equator is captured by the Cartesian simulations.

Radial profiles of the horizontal stress, $\tilde Q_{yz}$, from the
Cartesian simulations are shown in the middle panels of
Fig.~\ref{fig:cRss}, and latitudinal profiles in
Fig.~\ref{fig:pline_QhorOm_A5} with open squares and diamonds.
Similarly as in the spherical runs, this component peaks both at top
and bottom of the convective layer. However, some discrepancies are
observed between the profiles in different geometries. For instance,
in spherical Run~A4 the stress is somewhat more widely distributed
than in the corresponding Cartesian runs. In spherical Run~A5 the
radial profile of this component exhibits a bump at the bottom of the
convection zone which is much larger than in the corresponding
Cartesian cases.  Note, however, that in Fig.~\ref{fig:cRss}, the
uppermost peak moves inwards with increasing rotation between Sets~cA
and cD, and at the same time as the lowermost peak increases in
amplitude.  For the spherical Run~A6 with the highest Coriolis number
of roughly six, the stress changes sign in the region near the
surface, which is not visible in the Cartesian simulations with
Coriolis numbers of roughly four (Runs~cE1--cE4).

Finally, the meridional Reynolds stress, $\tilde Q_{xy}$,
corresponding to $\tilde Q_{r\theta}$, is positive
in the entire convection zone for moderate rotation (Runs~cA1--cA4). 
For larger
$\Co$, $\tilde Q_{xy}$ is negative in the lower part of the
domain (see the right panels of Fig.~\ref{fig:cRss}). 
Similar behaviour occurs in the spherical case with intermediate
rotation (Runs~A3--A5). In the most rapidly rotating case (Run~A6)
another sign change occurs near the equator (see
Fig.~\ref{fig:pRxyA}), which is not observed in Cartesian runs. 
This, however, could again be explained by the smaller $\Co$ in the
Cartesian runs.

\subsubsection{Filtering banana cells}
\label{sec:ban}
The large amplitude of the horizontal Reynolds stress,
peaking around $\pm7\degr$ latitude, has been an intriguing issue for
several years (e.g., Chan \cite{C01}; Hupfer et al.\ \cite{HKS05,HKS06}). 
One factor that might be contributing to the Reynolds stress are the
large-scale banana cell-like flows that develop near the equator
(e.g., K\"apyl\"a et al.\ \cite{KKT04}; Chan \cite{C07}). Such flows
vary in the azimuthal ($z$) direction and can lead to overestimation
of the contribution of turbulence, especially if averaging is
performed over the azimuthal ($z$) direction. We explore this
possibility by filtering out the contribution coming from the
large-scale structures observed in the $yz$-plane (the so-called
banana cells observed in spherical simulations). The procedure used in
this analysis is described below.

We perform a Fourier decomposition of the horizontal
velocities and find out at which Fourier mode the contribution of the
large scales peaks in the spectra. We find that the maximum is usually 
situated at wavenumber
$q=2$. Next we remove this mode from the spectra and make an 
inverse Fourier transformation, thus obtaining the velocity
field without the contribution from the large-scale
motions. Finally, we compute $Q_{yz}$ from the filtered velocities. 

Horizontal stress $Q_{yz}$ computed from filtered velocity fields
for Runs~cD1--cD4 for different latitudes
at $r=0.9R$ are plotted with blue square symbols in
Fig.~\ref{fig:pline_QhorOm_A5}. The radial variation of $Q_{yz}$ 
at $7\degr$ for Run~cD2
is shown with blue square symbols in Fig.~\ref{fig:cRss}. It is
clear from these figures that the $q=2$ mode is the dominant 
contribution to $Q_{yz}$ near the surface and it also affects
significantly the secondary peak in deeper layers. Thus, a 
flatter profile in latitude with a reduced amplitude of the 
stress is obtained in comparison to the non-filtered values.
The maximum, however, still resides around $\pm7\degr$, which is at
odds with theory (e.g.\ R\"udiger \& Kitchatinov \cite{RK07}).

\begin{figure}%[t]
\centering
\includegraphics[width=\columnwidth]{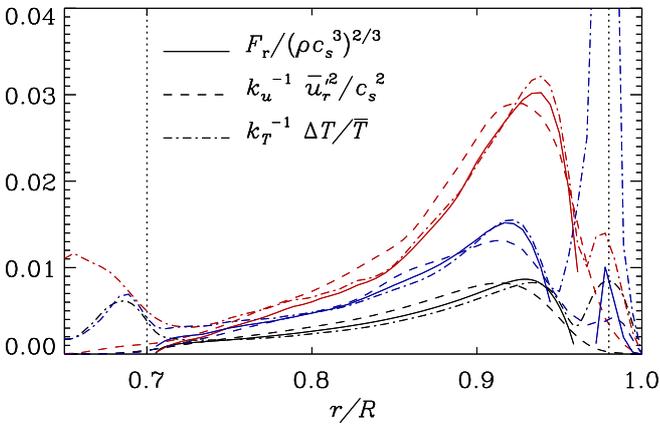}
\caption{Normalized radial turbulent heat flux raised to the 2/3 power
  as a function of $r$ ($x$) (solid lines). The dashed and dot-dashed
  lines correspond to the squares of the radial velocity and
  temperature fluctuations scaled with the coefficients $k_u$ and
  $k_T$, respectively.
  The upper (red), middle (blue) and lower (black) curves correspond 
  to Runs~cA0, cF0 and A0, respectively.
}
\label{fig:mlt}
\end{figure}

\subsection{Turbulent heat transport}
\label{sec:tuhe}

In non-rotating convection the radial heat flux,
\begin{equation}
F_r=c_{\rm P} \mean{\rho}\mean{u_r' T'}, 
\label{equ:cFlux}
\end{equation}
transports all of the energy through the
convection zone.
According to mixing length theory, velocity and temperature fluctuations
are related via $\mean{u_r'^2} \sim (\Delta T/\mean{T}) g \ell$,
where $\ell$ is the mixing length, $g \ell = \cst$, and $\Delta T =
\sqrt{\mean{T'^2}}$. Thus, the
three quantities are related via:
\begin{equation}
\frac{\Delta T}{\mean{T}} \sim \frac{\mean{u_r'^2}}{c_{\rm s}^2} \sim \left(
  \frac{F_r}{\rho c_{\rm s}^3} \right)^{2/3}.
\end{equation}
These quantities are shown in Fig.~\ref{fig:mlt} for non-rotating
simulations in Cartesian (Run~cA0) and spherical (Run~A0)
geometries. Here we use the coefficients
\begin{equation}
k_u = \frac{\langle \mean{u_r'^2} /c_{\rm s}^2 \rangle_{\rm CZ}} {\langle
  F_r / \rho c_{\rm s}^3 \rangle^{2/3}_{\rm CZ}} ~, \quad 
k_T = \frac{\langle \Delta T / \mean{T} \rangle_{\rm CZ}} {\langle F_r /
  \rho c_{\rm s}^3 \rangle^{2/3}_{\rm CZ}} ~,
\label{equ:koot}
\end{equation}
where $\brac{.}_{\rm CZ}$ denotes an average over the convection
zone. For both geometries we obtain $k_u \approx 0.4$ and $k_T
\approx 1.3$, values that are in good agreement with previous
results (Brandenburg et al.\ \cite{BCNS05}). Note, however, 
that the magnitude of the flux in Cartesian coordinates is around four
times larger than that in the spherical one, implying a difference of 
$4^{1/3}\approx1.6$ in the radial velocities according to 
Eq.~(\ref{equ:koot}). This is 
roughly the same factor
seen in the rms velocities
(compare Runs~A0 and cA0). 
This difference 
arises from the fact that we are considering a depth dependent gravity
also in the Cartesian simulations.
In spherical geometry, the
luminosity is constant and the flux decreases outwards proportional to
$r^{-2}$, whereas in Cartesian geometry the flux is constant. This
means that for the same profile of thermal conductivity, a significantly
larger portion of the energy is transported by convection in the
Cartesian case. 
We verify this result with a separate Cartesian model in which 
the radiative flux is constant and, like in the other models, the
gravity varies with depth.  In this case the thermal conductivity 
varies with radius. 
The profiles of the quantities depicted in Fig.~\ref{fig:mlt} obtained from 
this run (see blue lines and Run~cF0 in Table~\ref{tab:car_runs}) are in 
better agreement with the spherical
case. Similar results have been obtained if both, radiative 
flux and gravity, are constant (Run~cF1).

\begin{figure}[t]
\centering
\includegraphics[width=.5\textwidth]{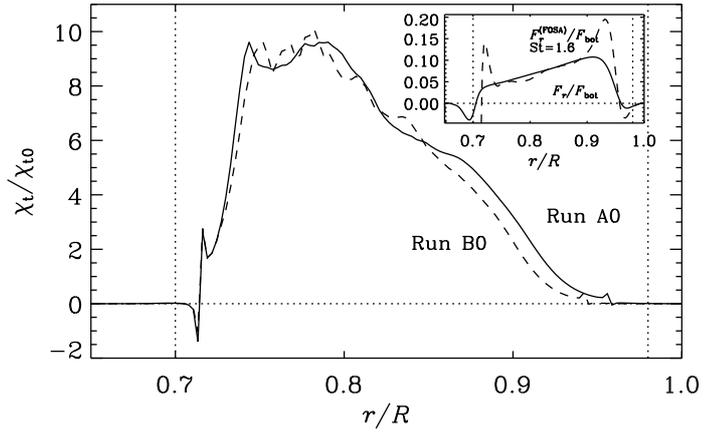}
\caption{Turbulent heat conductivity $\chit$ from Runs~A0 (solid line) 
  and B0 (dashed line). The inset
  shows the radial heat flux $F_r$ (solid line) and an analytical
  expression given in Eq.~(\ref{equ:Frfosa}) (dashed line) normalized
  by the heat flux at $r_1$ from Run~A0.}
\label{fig:pchit}
\end{figure}

The radial turbulent heat transport may also be described
in terms of a turbulent heat conductivity (e.g.\ R\"udiger
\cite{R89}) 
\begin{equation}
F_r=c_{\rm P} \mean{\rho}\mean{u_r' T'} \equiv -\mean{\rho}\mean{T}\chit \nabla_r \mean{s},
\label{equ:Flux}
\end{equation}
from which we can solve the turbulent heat conductivity as
\begin{equation}
\chit=-\frac{c_{\rm P} \mean{u_r' T'}}{\mean{T}\nabla_r \mean{s}}.
\label{equ:chit}
\end{equation}
The result, normalized by a reference value $\chitz=\urms/(3\kef)$,
for Runs~A0 and B0 are shown in Fig.~\ref{fig:pchit}. Here averages over
longitude and latitude are considered. We find that the value of
$\chit$ is almost ten times the reference value. The apparently large
value is most likely due to the normalization factor which is based on
a volume average of the rms velocity and a more or less arbitrary
length scale $\kef^{-1}$ (see also K\"apyl\"a et al.\
\cite{KBKSN10}). The sharp peaks and negative values of $\chit$
towards the bottom and top of the convectively unstable region reflect
the sign change of the entropy gradient which is not captured by
Eq.~(\ref{equ:chit}).

\begin{figure}[t]
\centering
\includegraphics[width=.5\textwidth]{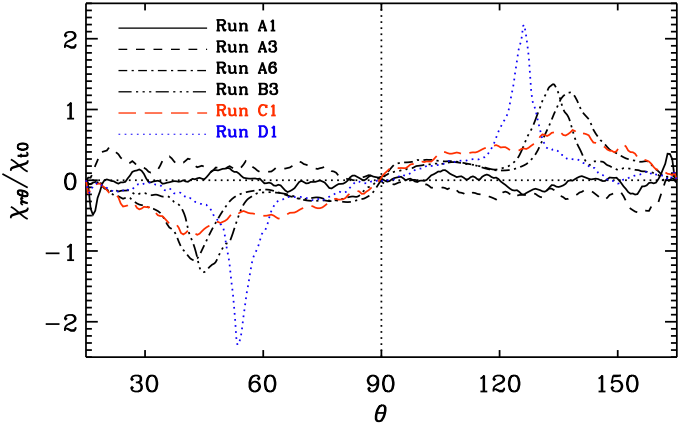}
\caption{Off-diagonal component $\chi_{\theta r}$ of the turbulent
  heat conductivity according to Eq.~(\ref{equ:chitr}) from Runs~A1
  (solid line), A3 (dashed), A6 (dot-dashed), B3 (triple-dot-dashed),
  C1 (red dashed), and D1 (blue dotted).}
\label{fig:pchitrA}
\end{figure}

\begin{figure}[t]
\centering
\includegraphics[width=\columnwidth]{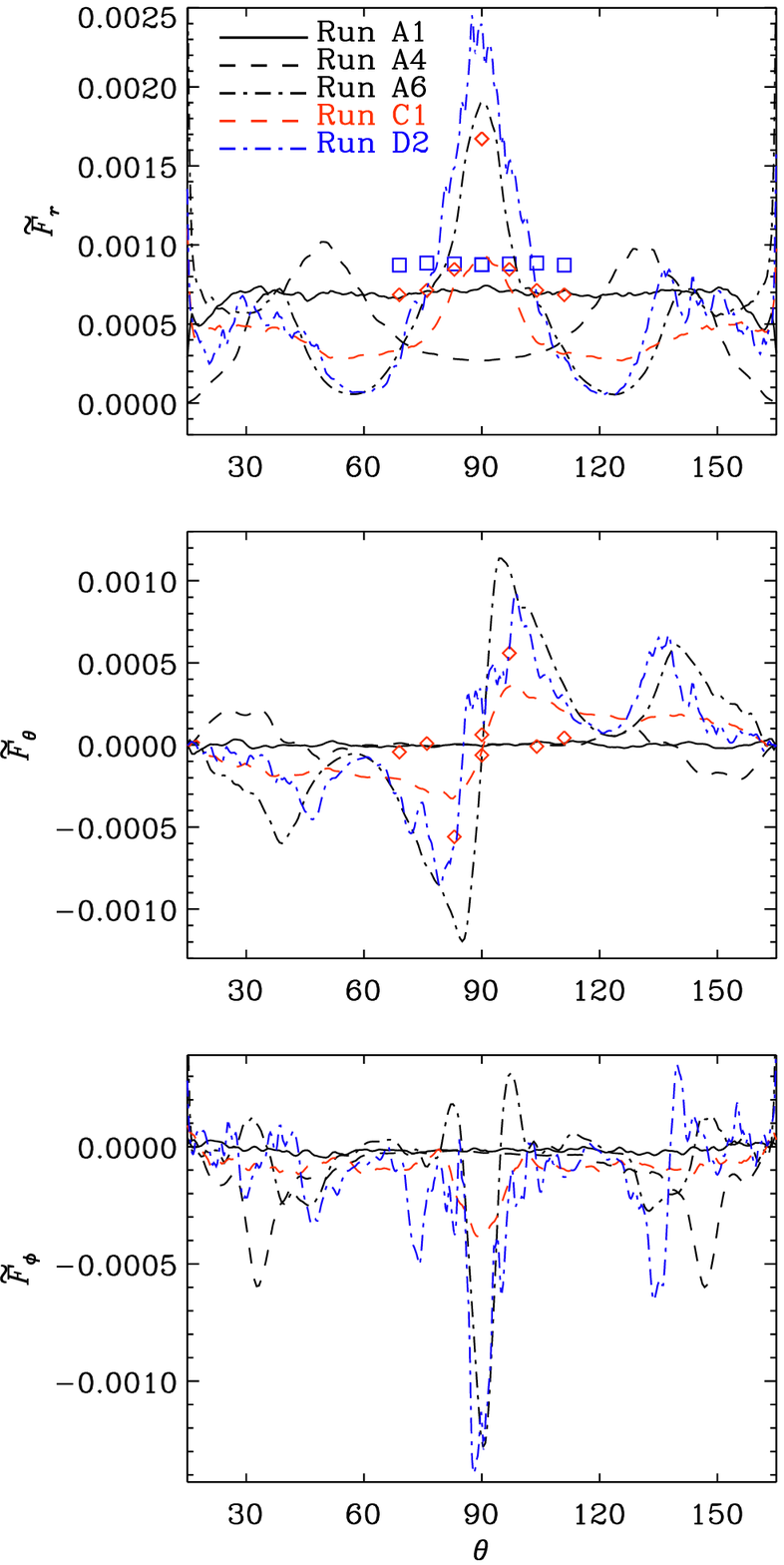}
\caption{Turbulent heat fluxes $\tilde{F}_r$ (top panel),
  $\tilde{F}_\theta$ (middle), and $\tilde{F}_\phi$ (bottom) from the
  runs indicated in the legend in the top panel. The symbols included
  in the top and middle panels correspond to vertical and latitudinal
  fluxes from Runs~cA1--cA4 (blue squares) and cE1--cE4 (red diamonds)
  scaled down by a factor of four (see the text for details). The
  data for Runs~C1 and D2 are scaled up by a factor of four.}
\label{fig:pfluxesA}
\end{figure}

According to first-order smoothing (e.g.\ R\"udiger \cite{R89}), the
radial flux can be written as
\begin{equation}
F_r^{\rm (FOSA)}=-\tau_{\rm c} \mean{u_r^2}~\mean{\rho}~\mean{T} \nabla_r \mean{s},
\label{equ:Frfosa}
\end{equation}
where $\tau_{\rm c}$ is the correlation time of turbulence. We compare
the actual radial heat flux with the rhs of Eq.~(\ref{equ:Frfosa})
in the inset of Fig.~\ref{fig:pchit}, where $\tauc$ is used as a 
fit parameter. A
reasonable fit within the convection zone is obtained if the Strouhal
number
\begin{equation}
\St=\tauc \urms \kef,
\end{equation}
is around 1.6 which is consistent with previous results from
convection (e.g.\ K\"apyl\"a et al.\ \cite{KBKSN10}). Note that the
ratio $\chit/\chitz$ gives a measure of the Strouhal number because
in the general case $\chitz=\onethird \tauc \urms^2=\St
\urms/(3\kef)$, whereas in the main panel of
Fig.~(\ref{fig:pchit}) we assume $\St=1$.

In rotating convection, Eq.~(\ref{equ:Flux}) no longer holds and the heat
flux becomes latitude-dependent. In mean-field theory this can be
represented in terms of an anisotropic turbulent heat conductivity
(Kitchatinov et al.\ \cite{KPR94})
\begin{equation}
\chi_{ij} = \chit \delta_{ij} + \chi_\Omega \varepsilon_{ijk} \hat{\Omega}_k + \chi_{\Omega\Omega} \hat{\Omega}_i \hat{\Omega}_j, 
\label{equ:Fluxt}
\end{equation}
where $\delta_{ij}$ and $\varepsilon_{ijk}$ are the Kronecker and
Levi--Civita tensors and $\hat{\Omega}_i$ is the unit vector along the
$i$th component of $\bm\Omega$.
This indicates that non-zero latitudinal and azimuthal heat fluxes are
also present in rotating convection. However, in order to compute all
relevant coefficients from Eq.~(\ref{equ:Fluxt}), a procedure similar
to the test scalar method (Brandenburg et al.\ \cite{BSV09})
would be required in spherical coordinates. In most of our runs,
however, the radial gradient of entropy is greater than the
latitudinal one. Thus we can approximate the latitudinal heat flux by
\begin{equation}
F_\theta = -\mean{\rho}\mean{T}\chi_{\theta r} \nabla_r \mean{s} -\mean{\rho}\mean{T}\chi_{\theta\theta} \nabla_\theta \mean{s}\approx -\mean{\rho}\mean{T}\chi_{\theta r} \nabla_r \mean{s},
\label{equ:chitr}
\end{equation}
from which the off-diagonal component $\chi_{\theta r}$ can be
computed in analogy to Eq.~(\ref{equ:chit}). Note that the sign of
$\chi_{\theta r}$ gives the direction of the latitudinal heat flux so
that positive (negative) values indicate equatorward (poleward) in the
northern (southern) hemisphere. According to Eqs.~(\ref{equ:Fluxt})
and (\ref{equ:chitr}), $F_\theta\propto \sin\theta \cos \theta$,
indicating a sign change at the equator.

\begin{figure*}[t]
\centering
\includegraphics[width=\textwidth]{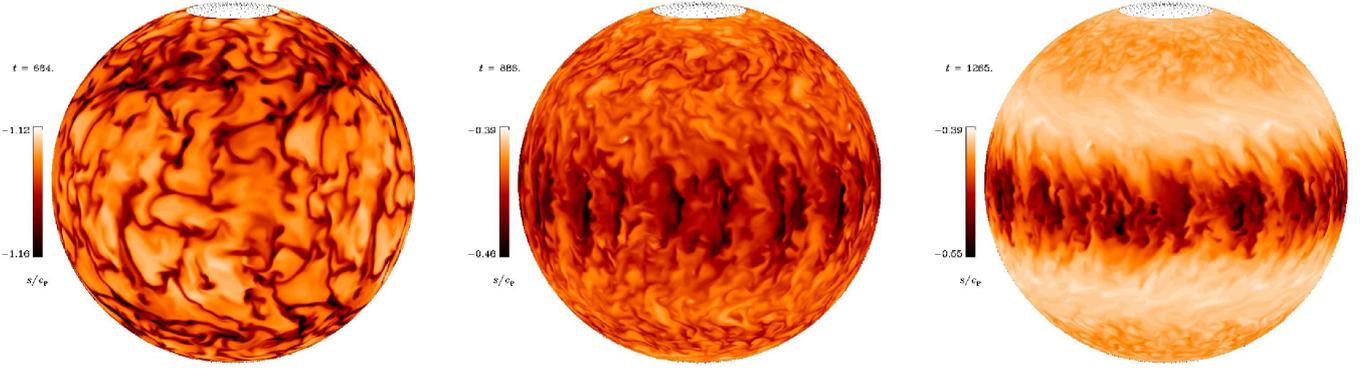}
\caption{Specific entropy in the upper part of the convectively
  unstable layer in Runs~B1 (left), D1 (middle), and D2 (right). The
  $\phi$-extent is duplicated fourfold for visualization purposes.}
\label{fig:eballs}
\end{figure*}

\begin{figure*}[t]
\centering
\includegraphics[width=.9\textwidth]{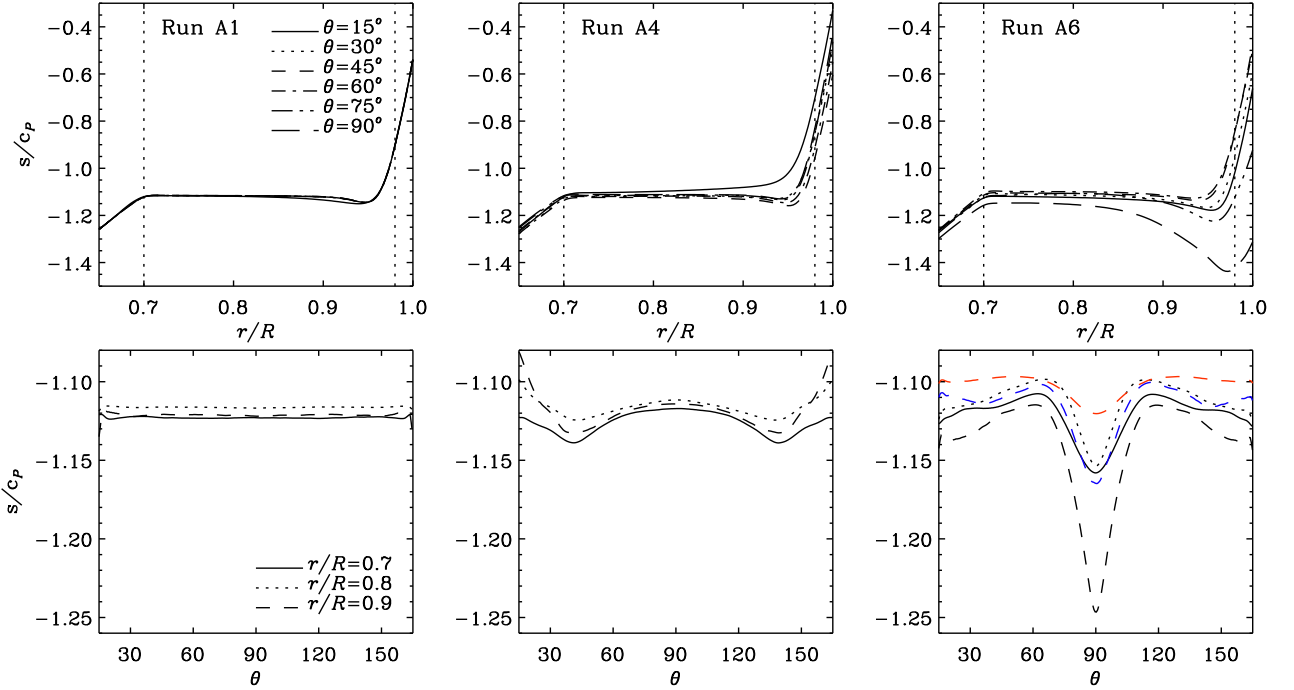}
\caption{Top row: radial profiles of entropy from six colatitudes as
  indicated by the legend in the leftmost panel from Runs~A1 (left
  column), A4 (middle column), and A6 (right column). Bottom row:
  latitudinal entropy profiles for the same runs as in the upper row
  at three radial positions indicated by the legend in the left
  panel. The red and blue dashed curves in the lower right panel show
  data at $r=0.9R$ from Runs~C1 and D2, respectively.}
\label{fig:pentA}
\end{figure*}

Representative results from Runs~A1, A3, A6, B3, C1, and D1 are shown in
Fig.~(\ref{fig:pchitrA}). For slow rotation (Run~A1), $\chi_{\theta
  r}$ is small and shows no coherent latitude dependence. In the
intermediate rotation regime (Run~A3), $\chi_{\theta r}$ is positive
(negative) in the northern (southern) hemisphere. In the most rapidly
rotating case (Runs~A6 and B3), the sign changes so that the heat flux is
towards the poles. 
Qualitatively similar results are obtained from rapidly rotating
Runs~C1 and D1 with a lower Mach number. The smoother latitude profile of 
$\chi_{r \theta}$ in Run~C1 reflects the smoother entropy profile 
(see Fig.~\ref{fig:pentA}).
The qualitative behaviour as a function of rotation is similar to that found
in local simulations (K\"apyl\"a et al.\ \cite{KKT04}). Comparing with
Fig.~\ref{fig:pchit} we find $\chi_{\theta r}/ \chit \equiv
\chi_{\theta r}/\chi_{rr}\approx 0.1$, which is of the same order of
magnitude as in local convection models K\"apyl\"a et al.\
(\cite{KKT04}) and forced turbulence Brandenburg et al.\
(\cite{BSV09}).
We note that the latitudinal entropy gradient, which we neglected 
in Eq.~(\ref{equ:chitr}), can become comparable
with the radial one in the rapid rotation regime near the equator.
Since $\nabla_\theta \mean{s}<0$ in the northern hemisphere (cf.\ 
Fig.~\ref{fig:pentA}), the latter term in Eq.~(\ref{equ:chitr}) 
yields a positive contribution to the flux. Thus
our values of $\chi_{\theta r}$ near the equator are likely to be
underestimated in the rapid rotation regime.
We postpone a more
detailed study of the turbulent transport coefficients to a future
publication and discuss the different components of the turbulent heat
fluxes.
We present the components of
convective energy flux as
\begin{equation}
\tilde{F}_i=F_i/\mean{\rho}\, \mean{c_{\rm s}}^3,
\end{equation}
where longitudinal averages are used.

Figure~\ref{fig:pfluxesA} shows the normalized turbulent heat fluxes as 
functions
of latitude from five runs with slow (Run~A1), intermediate (Run~A4),
and rapid (Runs~A6, C1, and D2) rotation.
We find that $\tilde{F}_r$ shows
little latitudinal variation except near the latitudinal boundaries
for slow and moderate rotation (Runs~A1--A3). For intermediate
rotation $F_r$ peaks at mid latitudes (Runs~A4--A5) whereas in the most
rapidly rotating cases (Runs~A6, C1, and D2) the maxima occur 
near the equator and at the
latitudinal boundaries. This behaviour follows the trend seen in the
entropy profile (Fig~\ref{fig:pentA}): the radial gradient of entropy
shows only a minor variation as a function of latitude in the most
slowly rotating runs (A1--A3). In Runs~A4 and A5 the gradient is
the steepest at mid latitudes and at the equator in Run~A6.
We find that the entropy gradient can become positive at certain
latitudes, e.g.\ close to the pole for Run~A4 and around latitudes
$\pm30\degr$ in Run~A6.

The horizontal fluxes, $F_\theta$ and $F_\phi$ are negligibly small in
comparison to the radial flux $F_r$ in the slow rotation regime
(Run~A1). The latitudinal flux is consistent with zero for all depths in
Run~A1 (see Fig.~\ref{fig:pfluxesA}).
For intermediate rotation (Runs~A2--A4) the latitudinal flux is mostly
equatorward. For the most rapidly rotating cases the sign changes so
that in Runs~A6, C1, and D2, $\tilde{F}_y$ is mostly poleward 
in the convection
zone. The magnitude of the latitudinal flux also increases so that the
maximum values, which are located near the surface, can become comparable 
with the radial flux.
The azimuthal flux is also small
and always negative, i.e., in the retrograde or westward direction, in
accordance with the results of R\"udiger et al.\ (\cite{R05a}) 
and Brandenburg et al.\ (\cite{BSV09}).

In some of the panels in Fig.~\ref{fig:pfluxesA} we also present 
results from Cartesian simulations (see the red and blue
symbols) from the same depth. As discussed above, the fluxes are
larger in this geometry, due to which we have scaled the
fluxes down by a factor of four in this figure.
We find that the latitude profiles of the radial and latitudinal heat
fluxes in the Cartesian simulations are in rather good agreement with
the spherical results. This is more clear in the rapidly rotating
cases cE1--cE4 in comparison to Run~A6 (see the right panels of
Fig.~\ref{fig:pfluxesA}), where the large peak of $F_r$ at the
equator, and the sharp peak of $F_\theta$ at low latitudes are
reproduced.

We find that the latitudinal entropy profiles show a local maximum
(slow and intermediate rotation) or a minimum (rapid rotation) at the
equator, see Fig.~\ref{fig:eballs} and the bottom panels
of Fig.~\ref{fig:pentA}.
The entropy profiles in the most rapidly rotating simulations (Run~A6 and B3)
are similar to that obtained by Miesch et al.\ (\cite{Mea00}) but
differs from the more monotonic profiles of e.g.\ Brun et al.\
(\cite{BT02}) and the lower Mach number case Run~C1.

\begin{figure}%[t]
\centering
\includegraphics[width=\columnwidth]{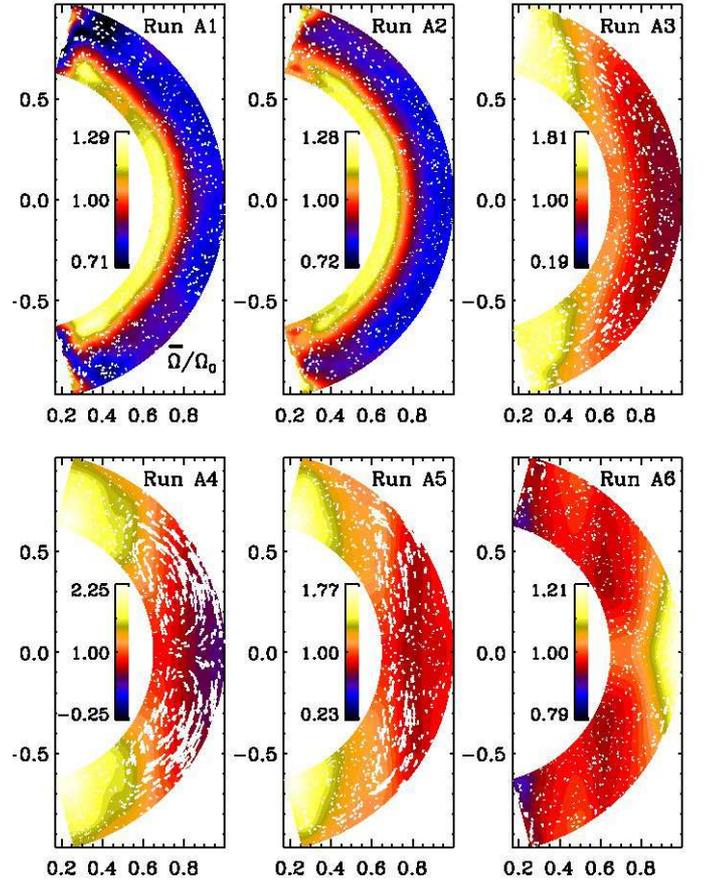}
\caption{Azimuthally averaged flows from the runs in Set~A. The contours show
  $\mean{\Omega}=\mean{u}_\phi/(r \sin\theta) + \Omega_0$ and the
  white arrows denote the meridional circulation.}
\label{fig:pOmA}
\end{figure}

\subsection{Large-scale flows}
\label{sec:LS}
The rotation profiles from the runs in Set~A are shown in
Fig.~\ref{fig:pOmA}. For slow rotation (Runs~A1--A2), a clear
large-scale radial shear, almost independent of
latitude, develops.
This is an old result going back to Kippenhahn (\cite{Kip63})
that is expected for turbulence whose vertical motions dominate over
horizontal ones (R\"udiger \cite{R89}).
Such a result has been obtained
in many mean-field models (e.g.\ Brandenburg et al.\ \cite{BMRT90}) 
and simulations since then (Brun \& Palacios \cite{BP09},
who refer to such flows as shellular).
However, the $\Omega$--profiles in these runs are clearly
different at high latitudes, which is probably an artefact due to
the latitudinal boundaries. As the Coriolis number is increased, the
radial shear remains negative, equatorial deceleration grows, and the
isocontours of $\Omega$ tend to align more with the rotation vector 
(Runs~A3--A4) -- in accordance with the Taylor--Proudman theorem.
Similar anti-solar rotation profiles have been reported also by
Rieutord et al.\ (\cite{RBMD94}), Dobler et al.\ (\cite{DSB06}), 
Brown (\cite{B09}), and Chan (\cite{C10}).
Such rotation profiles are usually the result of strong meridional
circulation (Kitchatinov \& R\"udiger \cite{KR04}) which is consistent 
with the present results.
Run~A5
represents a transitory case where bands of faster and slower rotation
appear, whereas in Run~A6 a solar-like equatorial acceleration
is seen. 
Similar transitory profiles have recently been reported by Chan 
(\cite{C10}).
The rotation profile in Run~A6 is dominated by the
Taylor--Proudman balance and the latitudinal shear is concentrated in
a latitude strip of $\pm30\degr$ about the equator.
Similar $\Omega$--profiles have been obtained earlier from more
specifically solar-like simulations (e.g., Brun \& Toomre \cite{BT02};
Brun et al.\ \cite{BMT04}; Brown et al.\ \cite{BBBMT08}; 
Ghizaru et al.\ \cite{GCS10}).

In the slow rotation regime (Runs~A1--A2) the kinetic energy of
meridional circulation and differential rotation are comparable and comprise a
few per cent of the total kinetic energy (columns 9 and 10 in
Table~\ref{tab:runs}). Increasing the Coriolis number further,
increases the fraction of kinetic energy in the differential rotation
whereas that of the meridional circulation remains at first constant
(Runs~A3--A4), and finally drops close to zero (Runs~A5--A6). In the
three most rapidly rotating cases the differential rotation comprises
more than 80 per cent of the total kinetic energy. We also find that the
meridional circulation shows a coherent pattern {\em only} for intermediate
rotation rates (Runs~A3--A5) where a single counter-clockwise cell per
hemisphere appears. In Run~A6 the meridional flow is concentrated in a
number of small cells in accordance with earlier results (e.g.,
Miesch et al.\ \cite{Mea00}; Brun \& Toomre \cite{BT02}).
We note that the rotation profiles in Runs~B3, C1, and D2 are similar to
that in Run~A6.

The surface differential rotation of stars can be observationally
studied using photometric time series (e.g.\ Hall \cite{Hall91}) or
with Doppler imaging methods (for a review, see Collier-Cameron
\cite{CC07}). The amount of surface differential rotation has been
determined for some rapidly rotating pre- or main-sequence stars with
varying spectral type (F, G, K, and M), systematically showing
solar-type differential rotation pattern with a faster equator and
slower poles. The strength of the differential rotation shows a clear
trend as function of the effective temperature, the shear being
larger for hotter stars (see Fig.\ 1 of Collier-Cameron
\cite{CC07}). Analysis of photometric time series, interpreting the
period variations seen in the light curve analysis being due to
differential rotation (e.g.\ Hall \cite{Hall91}), have established a
relation $\Delta \Omega / \Omega_0 \approx \Omega^{-n}$, with the
values of $n \approx$ 0.8--0.9. The observational results are in rough
agreement with theoretical predictions (e.g.\ Kitchatinov \& R\"udiger
\cite{KR99}), the theory predicting slightly weaker differential
rotation in the rapid rotators than the actually observed values.

\begin{figure}[t]
\centering
\includegraphics[width=\columnwidth]{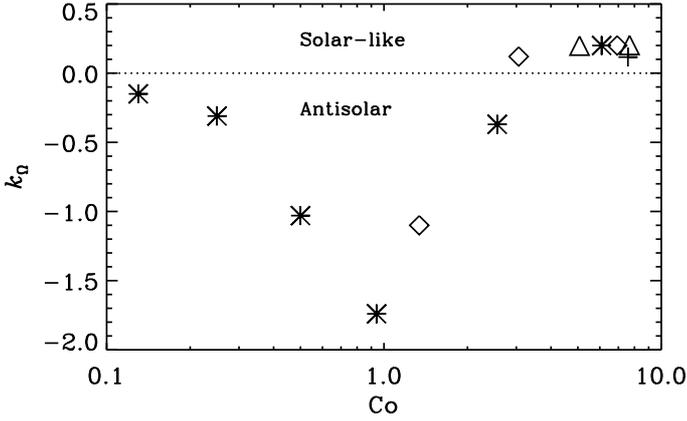}
\caption{Differential rotation parameter $k_\Omega$ according to
  Eq.~(\ref{equ:kO}) from Sets~A (stars), B (diamonds), Run~C1
  (cross), and Runs~D1 and D2 (triangles).
  The dotted horizontal
  line indicates the zero level.}
\label{fig:pk}
\end{figure}

We parameterise the differential
rotation in our simulations with the quantity
\begin{equation}
k_\Omega\equiv\frac{\Omega_{\rm eq}-\Omega_{\rm pole}}{\Omega_{\rm eq}}=\frac{\Delta \Omega}{\Omega_{\rm eq}},
\label{equ:kO}
\end{equation}
where $\Omega_{\rm eq}=\mean{\Omega}(r_4,\theta=90\degr)$ and
$\Omega_{\rm pole}=\mean{\Omega}(r_4,\theta=\theta_0)$. The results
for the runs with $\Co\neq0$ listed in Table~\ref{tab:runs} are shown
in Fig.~\ref{fig:pk}. We find that the anti-solar differential rotation
peaks at $\Co\approx1$ and that $k_\Omega$ turns positive for roughly
$\Co\approx3$. The values in the rapid rotation ($k_\Omega\approx0.2$)
end are comparable with the Sun (see also Chan \cite{C10}). It is not
clear, however, how realistic it is to compare the current simulations
with observations, i.e.\ even to argue that slowly rotating stars have
anti-solar differential rotation. It is clear that in the Sun the
Coriolis number, and the radial length scale of convection, vary much
more than in the current models so that it is not possible to
reproduce equatorial acceleration and surface shear layer
self-consistently in a single simulation. The situation may be
different in slow rotators but observing their differential rotation
is much more difficult. However, investigating the scaling of
$k_\Omega$ in the rapid rotation regime is likely worth pursuing
(see also Brown et al.\ \cite{BBBMT08}).

\section{Conclusions}

The present results have demonstrated that the basic properties of 
Reynolds stress and turbulent heat flux found in Cartesian simulations
are reproduced by simulations in spherical shells and wedges.
This includes the signs of the off-diagonal components
of $Q_{ij}$.
In particular, the vertical stress, $Q_{r\phi}$,
is negative in both hemispheres when $\Co$ is small,
but becomes positive near the top (and possibly also deeper down) when
$\Co$ is large.
This trend is well reproduced by the Cartesian simulations
where $Q_{xz}$ is also negative for small $\Co$, but
becomes positive near the top
when $\Co$ is large.
These results coincide with earlier findings of K\"apyl\"a et al.\
(\cite{KKT04}), Chan (\cite{C01}), and Robinson \& Chan 
(\cite{RC01}).

The horizontal stress $Q_{\theta\phi}$, with the counterpart $Q_{yz}$ 
in the Cartesian model, is found to be positive
in the northern hemisphere and have local maxima near the top and bottom of
the domain.
In spherical runs $Q_{\theta\phi}$ is found to change sign near the poles
for intermediate rotation.
For rapid rotation, $Q_{yz}$ reaches a maximum
near the top (or surface) around
$\pm7^\circ$ latitude -- in agreement with earlier results
(e.g., Chan \cite{C01}; Hupfer et al.\ \cite{HKS05}).
We show that large-scale velocities due to the banana cells near the
equator are the main contribution to $Q_{yz}$ in Cartesian calculations.
The spherical simulations reproduce such a sharp peak in the regime
$\Co\gtrsim 1$, the peak being limited to a radially narrow region near the
bottom of the domain.
We find that the results for the Reynolds stress are weakly dependent
on the Reynolds and Mach numbers.

Furthermore, we find that $Q_{r\theta}$ is positive in the northern
hemisphere, although for large values of $\Co$ the sign changes at
the bottom of the convection zone. For the largest value of $\Co$,
$Q_{r\theta}$ is negative throughout the entire convection zone. A
similar trend is seen in the Cartesian simulations, where $Q_{xy}$ is
mostly positive but becomes negative near the bottom of the convection
zone when rotation becomes strong enough, in accordance with
K\"apyl\"a et al.\ (\cite{KKT04})

The radial heat flux shows a strong dependence on latitude only when
rotation is fairly rapid, i.e.\ $\Co\gtrsim 1$. This is associated
with regions of the convection zone where the radial entropy gradient is 
decreased or even becomes
positive. A partial explanation is that our setup
(with a polytropic index of $n=1$) is such that roughly
80 per cent of the energy is transported by radiative diffusion
(cf.\ Brandenburg et al.\ \cite{BCNS05}), making
convection more easily suppressed than in a system where convection
transports a larger fraction.
The latitudinal heat flux is equatorward for slow rotation and changes
sign around $\Co\approx1$. A poleward heat flux is often used in
breaking the Taylor--Proudman balance (e.g.\ Brandenburg et al.\
\cite{BMT92}). Longitudinal heat flux is mostly in the retrograde
direction irrespective of the rotation rate.

The turbulent heat conductivity $\chit$ is comparable to the first-order
smoothing estimate with Strouhal number of the order of unity. The
off-diagonal component $\chi_{\theta r}$ is typically an order of
magnitude smaller than the diagonal component $\chit$ in the rapid
rotation regime. Similar results have been obtained previously from
local convection simulations (e.g.\ Pulkkinen et al.\ \cite{Pea93})
and forced turbulence (Brandenburg et al.\ \cite{BSV09}). In
mean-field models where anisotropic heat transport is invoked to break
the Taylor--Proudman balance, the anisotropic part is typically of the
same order of magnitude as the isotropic contribution (e.g.\
Brandenburg et al.\ \cite{BMT92}). It is conceivable that the
anisotropic contribution increases when the fraction of convective
energy flux is increased. However, such a study is not within the scope
of the present paper.

As discussed in Sect.~\ref{sec:Rey}, the components of the Reynolds
stress have contributions from diffusive and non-diffusive components.
In future work we hope to be able to separate these two contributions,
but in order to compare with earlier work, we have restricted ourselves
to studying the components of the Reynolds stress directly.
By making reasonable assumptions about the turbulent viscosity,
it is indeed possible to obtain the relevant components of the
$\Lambda$-effect, as was done by Pulkkinen et al.\ (\cite{Pea93}).
This is also true of global models, which also yield directly the
global flow properties that can then be compared with corresponding
mean field models, as was first done by Rieutord et al.\ (\cite{RBMD94}).
In a steady state, the Reynolds stress from the mean flow must then
balance both the viscous stress and the Reynolds stress from the
fluctuations, as was demonstrated also by Miesch et al.\ (\cite{MBDRT08}).
Such results are, however, dependent on the particular properties of the model.

In the present paper
we find that in the slow and intermediate rotation regimes the
differential rotation is anti-solar: the equator is rotating slower
than the high latitudes. Such rotation profiles also coincide with the
occurrence of coherent meridional circulation that is concentrated in
a single counter-clockwise cell. In the rapid rotation regime, solar-like
equatorial acceleration is obtained, but the differential rotation is
confined to latitudes $\pm30\degr$ and the isocontours are aligned
with the rotation vector.

To reproduce the solar rotation profile at least two major
obstacles remain. Firstly, the Taylor--Proudman balance must be
broken. A possibility is to use subgrid-scale models where the present
results for anisotropic heat transport can work as a guide. Secondly,
the Coriolis number should decrease near the surface so that the
transport of angular momentum is inward near the surface, leading to a
surface shear layer as in the Sun. Here we can again introduce a
subgrid-scale Reynolds stress guided by the present results. Studying
such models, however, is postponed to future papers.

\begin{acknowledgements}
We thank Dhrubaditya Mitra for useful discussions and an anonymous
referee for critical comments on the paper.
  The computations were performed on the facilities hosted by CSC --
  IT Center for Science Ltd. in Espoo, Finland, who are administered
  by the Finnish Ministry of Education.
We also acknowledge the allocation of computing resources provided by the
Swedish National Allocations Committee at the Center for
Parallel Computers at the Royal Institute of Technology in
Stockholm and the National Supercomputer Centers in Link\"oping.
This work was supported in part by
Academy of Finland grants 121431, 136189 (PJK), and 112020 (MJK),
the European Research Council under the AstroDyn Research Project 227952
and the Swedish Research Council grant 621-2007-4064.
\end{acknowledgements}

\appendix

\section{Dependence on domain size}
\label{depsize}

\begin{table*}[t!]
\centering
\caption[]{Summary of the runs with varying $\Delta \theta$ and $\Delta \phi$.}
% The runs are found in the folder:
% \texttt{pencil-code/petri/convection/spherical/turbtra}
      \label{tab:runs2}
      \vspace{-0.5cm}
     $$
         \begin{array}{p{0.05\linewidth}ccccccccccccc}
           \hline
           \noalign{\smallskip}
Run & $grid$ & \theta_1 & \Delta\theta & \Delta\phi & \Ra   & \Ma & \Rey & \Co & E_{\rm ther} & E_{\rm kin} & E_{\rm mer}/E_{\rm kin} & E_{\rm rot}/E_{\rm kin} & \Delta\Omega/\Omega_{\rm eq}  \\ \hline 
A5 & 128\times256\times128 & 15\degr & 150\degr & 90\degr &3.1\cdot10^6 & 
0.022 & 36 & 2.56 & 0.111 & 9.9\cdot10^{-4} & 0.002 & 0.949 & -0.37 \\ % 128x256x128a55
\hline
E1 & 128\times256\times32 & 15\degr & 150\degr & 22.5\degr &3.1\cdot10^6 &
0.019 & 31 & 2.91 & 0.113 & 7.0\cdot10^{-4} & 0.009 & 0.963 & -0.43 \\ % 128x256x32a5
E2 & 128\times256\times64 & 15\degr & 150\degr & 45\degr &3.1\cdot10^6 &
0.020 & 33 & 2.77 & 0.112 & 7.7\cdot10^{-4} & 0.008 & 0.946 & -0.35 \\ % 128x256x64a5
E3 & 128\times256\times256 & 15\degr & 150\degr & 180\degr &3.1\cdot10^6 &
0.022 & 37 & 2.47 & 0.113 & 1.1\cdot10^{-3} & 0.002 & 0.941 & -0.41 \\ % 128x256x256a5
E4 & 128\times256\times384 & 15\degr & 150\degr & 270\degr &3.1\cdot10^6 &
0.023 & 39 & 2.36 & 0.111 & 8.3\cdot10^{-4} & 0.002 & 0.902 & -0.29 \\ % 128x256x384a5
E5 & 128\times256\times512 & 15\degr & 150\degr & 360\degr &3.1\cdot10^6 &
0.025 & 41 & 2.23 & 0.112 & 4.0\cdot10^{-4} & 0.001 & 0.659 & -0.05 \\ % 128x256x512a5
           \hline
F1 & 128\times96\times128 & 60\degr &  60\degr & 90\degr &3.1\cdot10^6 &
0.033 & 31 & 2.92 & 0.115 & 1.3\cdot10^{-4} & 0.001 & 0.644 & +0.12 \\ % 128x96x128a5
F2 & 128\times160\times128 & 45\degr & 90\degr & 90\degr &3.1\cdot10^6 &
0.020 & 32 & 2.82 & 0.114 & 4.7\cdot10^{-4} & 0.004 & 0.899 & -0.13 \\ % 128x160x128a5
F3 & 128\times192\times128 & 30\degr & 120\degr & 90\degr &3.1\cdot10^6 &
0.021 & 34 & 2.68 & 0.113 & 4.3\cdot10^{-4} & 0.004 & 0.885 & -0.08 \\ % 128x192x128a5
F4 & 128\times288\times128 & 5\degr & 170\degr & 90\degr &3.1\cdot10^6 &
0.020 & 34 & 2.71 & 0.113 & 3.7\cdot10^{-4} & 0.006 & 0.882 & -0.07 \\ % 128x288x128a5
           \hline
         \end{array}
     $$
\end{table*}

Above we have shown that we can recover many earlier results obtained
in full spherical shells with wedges that span 150\degr
in latitude and 90\degr\ in longitude. This gives at least a fourfold
advantage in terms of computation time in comparison to a full
shell.
However, it is important to study the range within which we can still
recover the same results as with larger wedges.
In order to study this we
perform two additional sets of runs that are listed in
Table~\ref{tab:runs2}. In Set~E we vary the longitudinal extent from
$22.5\degr$ to full $360\degr$, with $\Delta\theta=150\degr$ in all
models. In Set~F we keep the longitudinal extent fixed at
$\Delta\phi=90\degr$ and vary the latitudinal extent between $60\degr$
and $170\degr$. As our base
model we take Run~A5 with fairly rapid rotation and complicated
large-scale flows in the saturated state.

\begin{figure*}[h]
\centering
\includegraphics[width=0.8\textwidth]{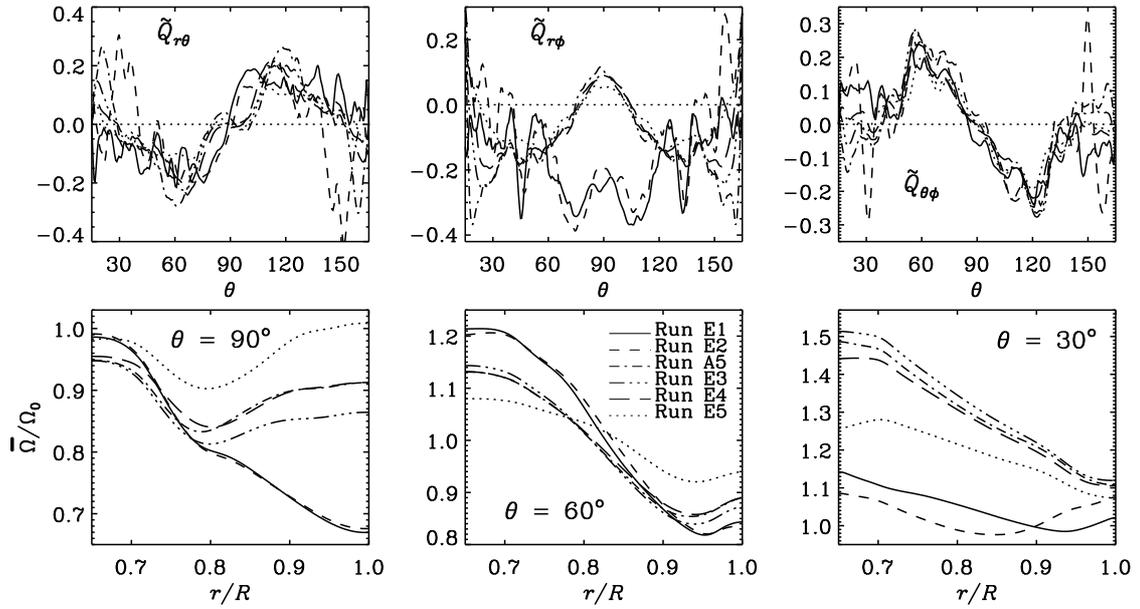}
\caption{Off-diagonal Reynolds stresses from the middle of the
  convection zone (upper row), and $\mean{\Omega}$ as a function of
  radius at $\theta=90\degr$ (lower row, left panel), $\theta=60\degr$
  (middle panel), and $\theta=30\degr$ (right panel) for Runs~E1--E5
  and A5. Linestyles as indicated in the legend in the lower middle
  panel.}
\label{fig:pOmE_line}
\end{figure*}

Figure~\ref{fig:pOmE_line} shows the latitudinal profiles of the
off-diagonal components of the Reynolds stress from the middle of the
convectively unstable layer and the rotation profiles as functions of
radius from three latitudes from Set~E and 
Run~A5. The Reynolds stresses are very similar in the latitude range 
$\pm45\degr$ in runs with
$\Delta\phi=90\degr$ or larger. Somewhat larger differences are seen
near the latitudinal boundaries. Runs~E1 and E2 with the
smallest longitude extents show the same qualitative behaviour for
stress components $\qrt$ and $\qtp$ but not for $\qrp$.
The rotation profiles for Runs~A5, E3, and E4 with
$\Delta\phi=90\degr-270\degr$ are very similar. The most obvious
deviations from the trend occur again for Runs~E1 and E2 where the
radial gradient of $\mean{\Omega}$ is negative at the equator as
opposed to the other runs where a positive gradient is found for
$r/R>0.8$. Surprisingly, Run~E5 with a full $360\degr$ longitude
extent also deviates from the trend seen in the intermediate
$\phi$-extents: the qualitative trend of $\mean{\Omega}$ is similar but
the magnitude of the differential rotation is reduced. This is due to
a non-axisymmetric $m=2$ mode which is excited in this
simulation. Large-scale hydrodynamical non-axisymmetries have been
reported from rapidly rotating convection (e.g.\ Brown et al.\
\cite{BBBMT08}). However, it is not clear whether the non-axisymmetry
in our Run~E5 is due to the same mechanism because of the slower
rotation.

\begin{figure*}[h]
\centering
\includegraphics[width=0.8\textwidth]{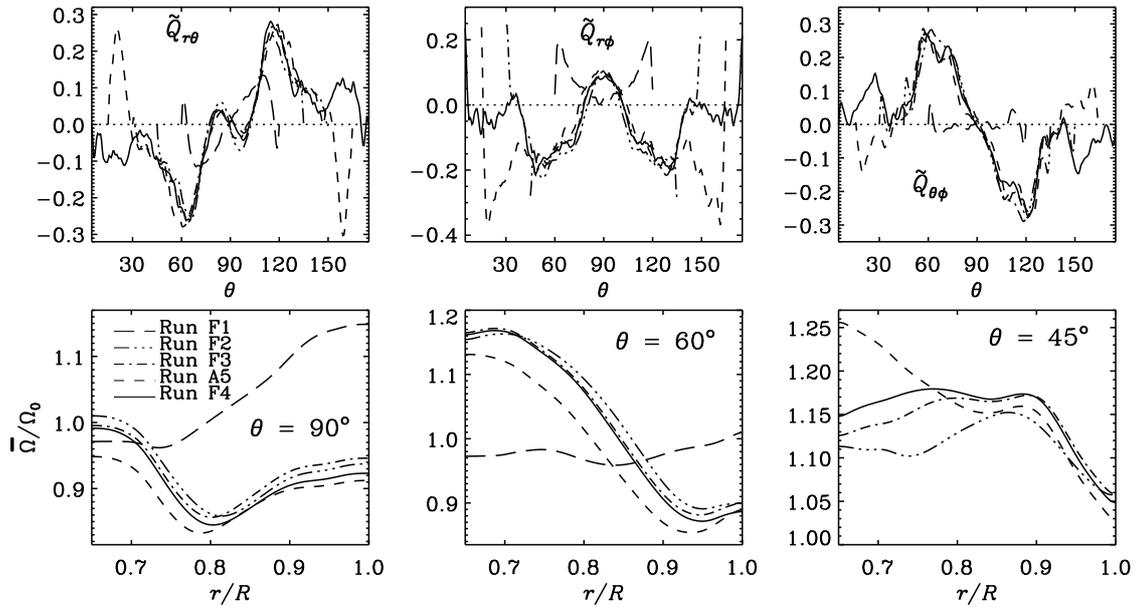}
\caption{Same as Fig.~\ref{fig:pOmE_line} but for Runs~F1--F4 and
  A5. The left panel on the lower row shows $\mean{\Omega}$ from
  $\theta=45\degr$. Linestyles as indicated in the legend in the lower
  left panel}
\label{fig:pOmF_line}
\end{figure*}

Comparing simulations with different latitudinal extents
(Fig.~\ref{fig:pOmF_line}), we find
that domains confined between $\pm45\degr$ latitude still reproduce
the essential features of the solutions. This is particularly clear
for the Reynolds stresses which are very similar in the latitude range
$\pm45\degr$ from the equator, with only Run~F1 showing qualitatively
different results in this range. There are also some differences at
high latitudes between Runs~A5 and F4. The
rotation profiles are also very similar in the range $\pm30\degr$
with the exception of Run~F1. Run~A5 also shows a deviating profile at
high latitudes.

These results suggest that a $90\degr$ longitude and $150\degr$
latitude extent is sufficient to capture the main features of the
solutions at larger domains. The cost of this is that some features
which are not of primary interest in the present study, such as the
large-scale non-axisymmetric modes, are omitted.

%\vspace{1cm} \noindent {\small \emph{$ $Id: paper.tex,v 1.321 2011-05-27 14:25:16 brandenb Exp $ $}

\end{document}